\documentclass[journal]{IEEEtran}

\usepackage{xcolor,soul,framed} 

\colorlet{shadecolor}{yellow}
\usepackage[pdftex]{graphicx}
\graphicspath{{../pdf/}{../jpeg/}}
\DeclareGraphicsExtensions{.pdf,.jpeg,.png}
\usepackage{bm}   
\usepackage[cmex10]{amsmath}
\usepackage{array}
\usepackage{float}
\usepackage[ruled]{algorithm2e}
\usepackage{mdwmath}
\usepackage{mdwtab}
\usepackage{pifont}
\usepackage[hidelinks]{hyperref}

\usepackage{url}
\usepackage{amssymb}
\usepackage{blindtext}
\usepackage{comment}
\usepackage{caption}
\usepackage{subcaption}
\usepackage{booktabs}
\usepackage{fancyhdr}
\begin{document}
\bstctlcite{IEEEexample:BSTcontrol}

\title{Transformer-Empowered Actor-Critic Reinforcement Learning for Sequence-Aware Service Function Chain Partitioning}

\author{
    Cyril Shih-Huan Hsu,
    Anestis~Dalgkitsis,
    Paola Grosso,
    Chrysa Papagianni

\thanks{University of Amsterdam, LAB42, Science Park 900, 1098 XH Amsterdam, The Netherlands (e-mail: s.h.hsu@uva.nl, a.dalgkitsis@uva.nl, c.papagianni@uva.nl, p.grosso@uva.nl).}}

\maketitle
\thispagestyle{fancy}

\begin{abstract}
In the forthcoming era of 6G networks, characterized by unprecedented data rates, ultra-low latency, and ubiquitous connectivity, effective management of Virtualized Network Functions (VNFs) is essential. VNFs are software-based counterparts of traditional hardware devices that facilitate flexible and scalable service provisioning. Service Function Chains (SFCs), structured as ordered sequences of VNFs, are pivotal in delivering complex network services. 
Nevertheless, splitting an SFC into multiple segments that are deployed across different network domains or infrastructure locations presents substantial challenges due to the potential heterogeneity of domain characteristic along with quality of service (QoS) constraints and limited visibility of network state.
Conventional optimization methods have limited scalability, while existing data-driven approaches struggle to balance efficiency with capturing VNF inter-dependencies in SFCs.
To overcome these limitations, we introduce a Transformer-empowered actor-critic framework specifically designed for sequence-aware SFC partitioning. By utilizing the self-attention mechanism, our approach effectively models complex inter-dependencies between VNFs, facilitating coordinated and parallel decision-making processes. Furthermore, to improve training stability and convergence we introduce an $\epsilon$-LoPe exploration strategy as well as Asymptotic Return Normalization. Comprehensive simulation results demonstrate that the proposed methodology outperforms existing state-of-the-art solutions in terms of long-term service acceptance rates, resource utilization, and scalability while achieving fast inference. 
\end{abstract}


\IEEEpeerreviewmaketitle

\section{Introduction}

\IEEEPARstart{T}{he} rapid emergence of beyond-5G and 6G networks has made network slicing and logical topology and resource abstraction central challenges in the optimization of network resources and services.
Network services compete for a limited pool of resources over the shared infrastructure to ensure robust Quality of Service (QoS) and meet stringent Service Level Agreements (SLA).
New services not only demand higher cross-domain performance guarantees but also operate in highly shared, multi-tenant environments, imposing significant pressure on both pre-deployment and post-deployment optimization processes.
In this context, service partitioning (i.e., distributing service functions across domains) becomes critical. By accounting for constraints, such as limited visibility of the network state, due to privacy regulations or domain-specific data governance policies, as well as controlled cost-driven overbooking, network operators can preserve performance and meet SLA commitments.

\begin{figure}[h!]
     \centering
     \includegraphics[width=1\columnwidth]{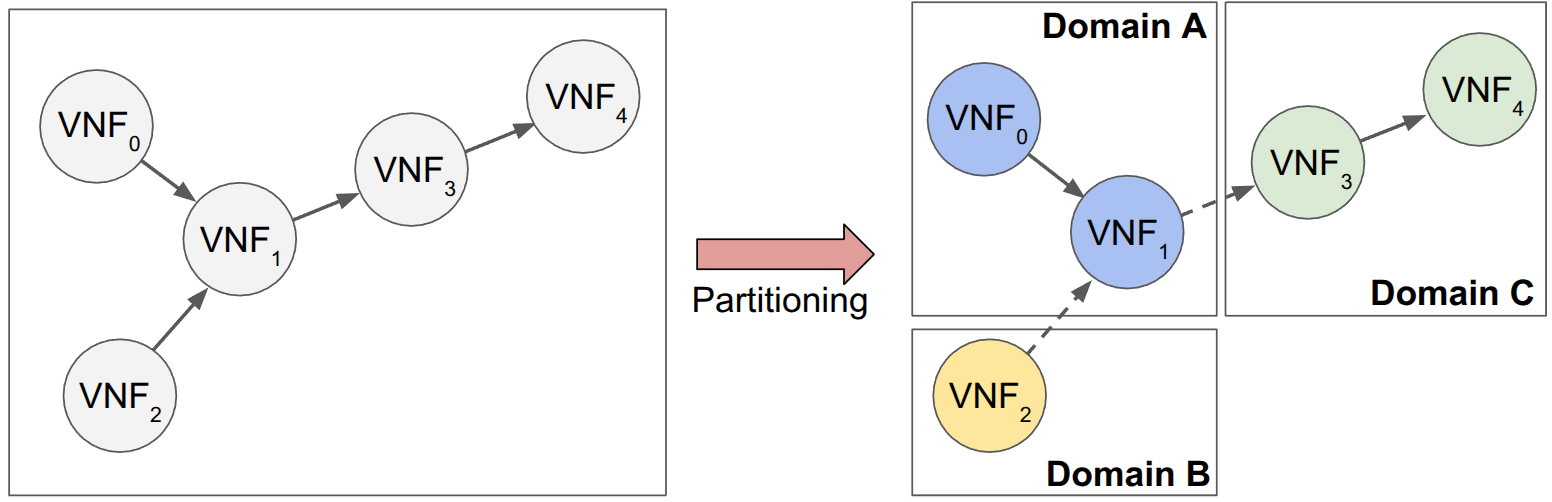}
    \caption{Illustration of SFC partitioning across multiple network domains. The left side presents a logical SFC composed of five interconnected VNFs. The right side shows the outcome of partitioning, where the VNFs are assigned to three separate domains: A, B, and C. Solid lines denote intra-domain connections, while dashed lines represent inter-domain links.}
    \label{fig:sfcp}
\end{figure}

The $3^{rd}$ Generation Partnership Project (3GPP) introduced End-to-End (E2E) logical network segmentation to enable per-service customization and management, powered by a range of technologies \cite{8423711}.
At its core, Software-Defined Networking (SDN) allows logically centralized, programmatic control of data flows.
ETSI has been standardizing Network Functions Virtualization (NFV) that decouples Network Functions (NF) from dedicated hardware and enables scaling services as network demands evolve \cite{BHAMARE2016138}.
Service Function Chaining (SFC) adds another layer of dynamism, allowing for the dynamic linking of multiple NFs to create flexible, policy-driven services. An SFC is essentially an ordered sequence of VNFs and is expressed in the form of a VNF-graph.
Fig.~\ref{fig:sfcp} illustrates the process of SFC partitioning across multiple network domains.


Orchestration platforms integrate these technologies, enabling the automation of network service lifecycle management, abstracting the complexities associated with creating, deploying, and reconfiguring network services and slices. This automation significantly reduces both Capital Expenditure (CAPEX) and Operational Expenditure (OPEX) by improving resource utilization, simplifying infrastructure management, and automating lifecycle operations, thereby lowering both infrastructure investment and ongoing maintenance costs for telecommunication operators.
However, the integration of heterogeneous domains, dynamic service requirements, and real-time decision-making introduces significant orchestration complexity.
Global service optimization algorithms have emerged to manage large-scale infrastructures involving multiple stakeholders and Infrastructure Service Providers (ISPs), raising privacy concerns among providers.
Service partitioning alleviates this by logically dividing an SFC into segments, facilitating separation of concerns, scaling, and partial orchestration without requiring full visibility into each provider's infrastructure.

Despite the advantages, Service Function Chain Partitioning (SFCP) poses a complex and challenging optimization problem.
The difficulty lies in jointly determining  (i) an optimal partitioning strategy and (ii) the placement of each segment across multiple network domains with heterogeneous and dynamic characteristics, including variations in resource capacity, capabilities, availability, and network dynamics. 
These domains, spanning terrestrial, aerial, satellite, and other network infrastructures, exhibit diverse behaviors that must be carefully accounted for in order to satisfy stringent QoS constraints.
This requires balancing resource availability, QoS requirements, inter-domain routing overhead, and data locality considerations,  often under incomplete or uncertain network state information. The combinatorial nature of possible partitioning and mapping decisions, coupled with dynamic network conditions and policy constraints, makes SFCP an NP-hard problem~\cite{9025750}, requiring scalable, intelligent, and privacy-preserving solutions.

Artificial Intelligence (AI) and Machine Learning (ML) have recently emerged as powerful tools for solving complex optimization tasks, often outperforming classical approaches in performance, efficiency, and cost-effectiveness \cite{LeCun2015, alphagozeroarticle}. Traditional optimization methods (e.g., approximation algorithms) and heuristics can yield near-optimal solutions, but they struggle to scale due to the combinatorial explosion of decisions in large infrastructures. To overcome this limitation, data-driven methods have been proposed, offering better scalability but introducing a fundamental trade-off. Sequential approaches model the SFC/VNF ordering and capture inter-dependencies between VNFs, yet suffer from slow inference, error propagation, and poor scalability. Parallel methods map all VNFs in one step, so they are faster and more scalable but often neglect coordination and ordering effects, leading to suboptimal results. Recent advances, including multi-agent Reinforcement Learning (RL) and Graph Neural Networks (GNN)-based approaches, attempt to balance these aspects but remain limited by coordination overhead, weak long-range dependency modeling, or training instability. Consequently, no existing method has achieved both scalability and true order-aware partitioning.

To this end, we investigate the problem of efficient SFC partitioning and propose \textbf{S}equence-aware \textbf{D}ifferentiable \textbf{A}ctor-\textbf{C}ritic
RL (SDAC), a novel approach that integrates Transformer layers into a differentiable actor-critic RL framework.
We evaluate the effectiveness of SDAC through an extensive and diverse simulation-based environment and compare it against multiple State-of-The-Art (SoTA) methods to showcase its ability to maintain a higher long-term service acceptance rate.
Thus, the contribution of this work is threefold:
\begin{itemize}
    \item We introduce a novel Deep-RL (DRL) framework  for partitioning ordered SFC graphs, which leverages Transformer encoders within an actor-critic architecture. The self-attention mechanism explicitly models inter-dependencies between VNFs, facilitating coordinated and parallel decision-making. Additionally, we introduce the $\epsilon$-LoPe exploration strategy and Asymptotic Return Normalization to enhance training stability and convergence. These innovations address the limitations of existing sequential and parallel VNF partitioning methods, significantly improving the efficiency and effectiveness of SFCP for multi-domain network service orchestration.
    \item We introduce an innovative critic network, inspired by sentence encoders used in Large Language Models (LLMs), that holistically evaluates the ordered sequence of VNF mapping decisions, providing a context-aware evaluation of the SFC partitioning strategy.
    By employing a continuous relaxation for discrete actions, the proposed framework enables E2E differentiable policy updates,  reducing reliance on stochastic sampling processes and mitigating the credit assignment problem.
    Its adaptability to a variable number of agents makes it a scalable solution for Multi-Agent Reinforcement Learning (MARL) problems, beyond the SFCP problem.
    \item We provide a comprehensive evaluation of SDAC through extensive simulations, demonstrating its superior performance compared to SoTA methods, including DRL- and (meta-)heuristic- based approaches. SDAC achieves higher long-term acceptance rates, improved resource utilization, and greater scalability while maintaining low inference times. These results validate the effectiveness of the proposed method for network management under varying conditions.
\end{itemize}
The remainder of this paper is organized as follows. Section~\ref{sec:related} offers a discussion of related work.
Section~\ref{sec:problem formulation} gives a comprehensive introduction to the problem formulation and system model.
Section~\ref{sec:proposed_method} details the proposed method.
Section~\ref{sec:perf_eval} describes the experimental setup.
Section~\ref{sec:results} presents an extensive performance evaluation study.
Finally, Section~\ref{sec:conclusion} concludes the paper and outlines future research directions. 
The source code is available via the \href{https://zenodo.org/communities/desire6g}{DESIRE6G community in Zenodo.}

\section{Related Work}
\label{sec:related}
The shift from hardware-coupled network architectures to flexible, software-driven ones has made service partitioning, along with related concepts such as SFCs, VNF Forwarding Graphs (FGs), and VNF-FG embedding, critical to system performance and reliability~\cite{BHAMARE2016138}.
The increasing complexity of multi-domain networks~\cite{8423711}, pushed recent literature efforts to integrate data-driven approaches into service partitioning strategies.
Table~\ref{tab:related_work_comparison} summarizes all approaches presented in this section. 
Specifically, we compare the approaches in terms of scalability, categorize the type of VNF coordination during SFC partitioning to sequential or parallel, and specify the decision horizon that the approach considers.
\begin{table*}[t]
\centering
\caption{Comparison of SFC Partitioning Approaches.}
\label{tab:related_work_comparison}
\begin{tabular}{|l|c|c|c|c|}
\hline
\multicolumn{1}{|c|}{\textbf{Approach}} & \textbf{Type} & \textbf{Scalability} & \textbf{Coordination} & \textbf{Decision Horizon}\\ \hline
(Mixed-)Integer programming (ILP, MILP)~\cite{SAHHAF2015492, 7820223, 9351537, 7145312} & Classical & Low & Parallel & Short-term \\ \hline 
(Meta-)Heuristics~\cite{6226390, 7994644, 7439948, MAGOULA2021108157, battisti2024novel} & Classical & Low & Parallel & Short-term \\ \hline 
Supervised Machine Learning (RNN, GNN)~\cite{9236954} & ML-based & Moderate & Sequential & Short-term \\ \hline
Single-Agent DRL (incl. GNN-based)~\cite{Santos2021, fake10107385, pei2019optimal, sun2020combining} & ML-based & Moderate & Sequential & Short-term \\ \hline
Multi-Agent DRL (incl. Federated RL)~\cite{9685162-SCHEMA, 9875038, fake10107385, wang2022multiagentdeepreinforcementlearning, kong2025dynamic} & ML-based & High & Sequential & Short-term \\ \hline
\textbf{Proposed Transformer-based DRL } & \textbf{ML-based} & \textbf{High} & \textbf{Parallel} & \textbf{Long-term} \\ \hline
\end{tabular}
\end{table*}

\subsection{Optimization-based Approaches}
Early work on SFC optimization largely relied on classical optimization formulations and heuristic algorithms, targeting multiple objectives. 
Initial studies addressed SFC partitioning, often jointly with embedding, through Integer Linear Programming (ILP) formulations. In~\cite{SAHHAF2015492}, the authors study service decomposition and propose an ILP formulation for minimizing mapping cost, complemented by a heuristic to address scalability. Dietrich et al. introduce Nestor~\cite{7145312}, a traffic-aware two-stage orchestration framework for service partitioning and intra-domain embedding across geographically distributed providers. This line of work is later extended in~\cite{7820223}, which formulates multi-provider service embedding as an ILP problem decomposed into SFC graph partitioning and mapping.
Abujoda et al.~\cite{7439948} propose \textit{DistNSE}, a distributed framework that partitions SFCs into sub-chains and embeds them without requiring operators to disclose internal topology or resource information. Battisti et al.~\cite{battisti2024novel} further study SFC segmentation in multi-domain topologies and propose a heuristic combinatorial strategy for producing valid segmentation plans while preserving the VNF order of linear chains.

Metaheuristic and game-theoretic formulations have also been explored for more dynamic and constrained settings. 
In~\cite{7994644}, the authors propose a game theory-inspired graph partitioning framework and prove the existence of a Nash equilibrium under server affinity, collocation, and latency constraints. 
 Magoula et al.~\cite{MAGOULA2021108157} study ultra-low-latency SFC placement and propose a delay- and location-aware Genetic Algorithm, accounting for processing, queuing, transmission, and propagation delays during VNF placement and path selection.  Li et al.~\cite{9351537} address dynamic SFC mapping and scheduling in Internet of Vehicles environments by formulating the problem as a Mixed-ILP (MILP) and solving it through two Tabu Search-based algorithms that account for VNF migration, reinstantiation, and rescheduling costs.

Although most of these approaches provide near-optimal solutions, they suffer from scalability limitations due to the exponential growth of possible actions. 
Specifically, each SFC may involve up to $M^N$ potential assignments, where $M$ is the number of available deployment targets and $N$ the number of VNFs in the SFC.
This combinatorial explosion, coupled with the lack of adaptability and long convergence times, substantially limits the practicality in real-world scenarios.

\subsection{Data-driven Approaches}
To overcome the limitations of classical optimization-based methods and embrace scalability, researchers pivoted their focus on AI data-driven solutions due to their remarkable performance that challenged the \textit{status quo} in networking.


RL-based solutions were introduced to address dynamic placement and orchestration. 
In~\cite{pei2019optimal}, Pei et al. formulate VNF placement as a Binary Integer Programming (BIP) problem under dynamic network load and combine it with Double Deep Q-Network (DDQN) learning to determine placement decisions. 
Sun et al.~\cite{sun2020combining} propose \textit{DeepOpt}, an RL and GNN-based approach for VNF placement that improves resource utilization and QoS under dynamic traffic and topology conditions. 
Similarly, Santos et al.~\cite{Santos2021} introduce an availability-aware SFC placement method based on Proximal Policy Optimization (PPO), while Heo et al.~\cite{9236954} propose a GNN-based encoder-decoder model that captures topology-aware node representations and supports placement in dynamically changing networks.

More recent works have explored multi-agent and federated learning schemes to address the scalability limitations of centralized solutions. 
Wang et al.~\cite{wang2022multiagentdeepreinforcementlearning} propose a cooperative MARL framework that decomposes VNF placement and routing into subtasks for heterogeneous requests. 
In the same direction, Pentelas et al.~\cite{fake10107385} propose a decentralized cooperative multi-agent RL scheme for SFC placement.
Kong et al.~\cite{kong2025dynamic} further extend this direction through \textit{MetaFedDRL}, an asynchronous federated RL framework for SFC deployment in heterogeneous multi-domain topologies, improving convergence, generalization, and horizontal scalability.
Sequential placement of functions offers stronger coordination but is time consuming and constrained by prior decisions, which can reduce overall efficiency~\cite{fake10107385, 9685162-SCHEMA, 9875038}. 

\subsection{Transformer-based Approaches}
Recently Transformers have emerged as a groundbreaking architecture that disrupted deep learning research by demonstrating outstanding results. 
The use of attention mechanisms enables the effective modeling of complex, long-range dependencies within data sequences. 
The authors in \cite{vaswani2017attention} propose the groundbreaking DNN architecture of Transformers and demonstrate the remarkable results in natural language translation.
Based on that, several studies \cite{khan2022transformers, brohan2022rt, islam2024comprehensive, yuan2025survey, shehzad2026graph} have showcased the successes of the architecture in various fields such as computer vision, natural language processing, audio processing and robotics.
This progress also motivated the integration of Transformers with DRL, where sequence modeling and temporal credit assignment are critical~\cite{yuan2024transformer, hu2024transforming, li2023survey, agarwal2023transformers}. 
A notable example is the Decision Transformer of Chen et al.~\cite{chen2021decision}, which reformulates RL as a conditional sequence modeling problem and demonstrates strong performance in long-horizon and delayed-reward settings.

In next-generation communication networks, Transformer-based methods have first been explored for physical-layer and radio optimization tasks. 
Specifically, the works in~\cite{wang2022transformer, raha2024advancing} discuss their strong representation capabilities in problems such as massive Multiple-Input Multiple-Output (MIMO) beamforming and semantic communications, while~\cite{9779340} applies Transformers to automatic modulation classification, improving accuracy at low Signal-to-Noise Ratios (SNRs) with fewer model parameters. 
Similarly,~\cite{9930825} proposes a temporal Transformer for real-time QoS prediction in heterogeneous Internet-of-Things (IoT) applications, utilizing the Attention mechanism to capture both short- and long-term traffic dependencies.
A temporal Transformer was proposed for real-time QoS prediction in heterogeneous Internet of Things (IoT) applications~\cite{9930825}, leveraging attention mechanisms for both short- and long-term sequence dependencies. Evaluated on diverse IoT traffic datasets, the model demonstrated robust and accurate QoS metric predictions.
Furthermore, a two-stage Vision Transformer (ViT) framework, MapViT, was introduced in~\cite{hsu2026mapvit} for real-time radio quality map estimation, showing high computational efficiency in dynamic environments.

Wu et al.~\cite{10147969} combine DDQN with a Decision Transformer for dynamic SFC placement, leveraging offline RL to improve generalization and reduce request rejections and E2E delay. Zheng et al.~\cite{zheng2025hra} propose \textit{HRA-SFCP}, integrating heuristics with a Graph Attention Transformer-based PPO agent for resource-aware SFC placement in LEO satellite networks.
Zhong et al.~\cite{zhong2024drl} introduce a Transformer-based DRL approach for reliability-aware SFC provisioning, while Wang et al.~\cite{wang2025distributed} develop a distributed generative RL framework using Decision Transformers and actor-critic learning for UAV swarm networks. 
In a different direction, Li et al.~\cite{li2025optimizing} combine Transformers with Ant Colony Optimization for cost-efficient routing and embedding, whereas Reilly et al.~\cite{reilly2026alpha} propose \textit{Alpha-SFC}, integrating a graph-based Transformer with Monte Carlo Tree Search for online SFC placement.
In~\cite{wu2024graph, habib2024transformer, chona2025transformer}, Transformer-based time-series models combine graph encoders to capture VNF dependencies and temporal load dynamics. Similarly, Song et al.~\cite{song2025deep} use a Transformer-enhanced PPO framework to schedule Directed Acyclic Graph (DAG) applications in edge networks, modeling task dependencies for optimized offloading and resource allocation.
While these studies highlight the strong potential of Transformers in 6G and networked systems, significant gaps remain in their integration into practical, real-time optimization, and orchestration for long decision horizons.

In summary, although data-driven methods improve scalability and long-term performance over traditional optimization ones, parallel decision-making often overlooks inter-VNF dependencies, while sequential decision-making captures them at the cost of computational efficiency. To address these gaps, we propose a Transformer-based actor-critic framework that models SFCs as token sequences, enabling parallel, context-aware, and scalable decision-making for efficient SFC partitioning.



\section{Problem Formulation}
\label{sec:problem formulation}

To effectively address the SFCP problem, we begin with a high-level overview, followed by a comprehensive description of the system model. This model serves as the foundation for capturing the interplay between the substrate network, SFC requests, and the constraints governing resource allocation. By formalizing these elements, we enable the optimization of long-term system performance while adhering to resource capacity and latency constraints.

\subsection{Problem Description}
An SFC is defined as an ordered sequence of VNFs that enforces a flow processing policy on network traffic. For example, a typical SFC might consist of VNFs such as \textit{VPN Gateway} $\rightarrow$ \textit{Firewall} $\rightarrow$ \textit{Encryption/Decryption} $\rightarrow$ \textit{Packet Inspection}, ensuring secure and efficient flow management for network applications.
Each VNF in the SFC requires specific computing resources (e.g., CPU) to process the network traffic, while the connections between consecutive VNFs, referred to as virtual links (vLinks), consume network bandwidth for traffic forwarding. These VNFs and vLinks are hosted on a multi-datacenter system, where Data Centers (DCs) offer computing resources via their servers, and the physical links within and between DCs provide the necessary bandwidth.


An SFC can be deployed either within a single domain or across multiple domains. In a single-domain setting, a single Infrastructure Provider (InP) manages the entire substrate network, enabling centralized control with full visibility of resource availability. In contrast, a multi-domain setting involves multiple administrative domains operated by different providers, that do not typically share internal state information. In this work, we focus on the multi-domain case, where a logically centralized Network Functions Virtualization Orchestrator (NFVO) performs inter-domain service orchestration in line with the ETSI NFV-MANO framework~\cite{etsi_mano}.


In this context \textit{partitioning} generally refers to the process of logically dividing an SFC into segments (sub-chains) and assigning them to  different domains, whereas \textit{embedding} refers to the subsequent mapping of these sub-chains onto physical resources. 
In this work, we refer to \textit{partitioning} as the mapping of VNFs at the inter-DC level, where DCs belong to different InPs, while \textit{embedding} refers to the fine-grained mapping of VNFs within a DC (intra-DC). The placement of vLinks is implicitly determined by the assignment of VNFs, as each vLink is mapped to a network path between the selected DCs. 

The SFCP problem is challenging due to multiple intertwined factors. VNFs and vLinks must be mapped under resource capacity, and E2E latency constraints to meet SLAs, while decision-making is complicated by limited visibility, as only aggregated resource information (e.g., total available CPU and bandwidth)  is typically available at the NFVO \cite{7820223}. The limited granularity of information available to the NFVO introduces significant uncertainty, as it cannot observe the exact state of the resource availability within each DC~\cite{fake10107385}. Moreover, strategies that optimize individual requests in isolation may lead to inefficient long-term resource utilization and fragmentation, ultimately reducing the system’s ability to admit future SFC requests and degrading the overall acceptance rate.
To summarize, in this study the objective of SFCP is to find a strategy that partitions SFCs so that ($i$) the resource constraints are satisfied; ($ii$) E2E latency requirement is met; and ($iii$) the long-term \textit{acceptance rate} is maximized.

\begin{figure}[h]
     \centering
     \includegraphics[width=1\columnwidth]{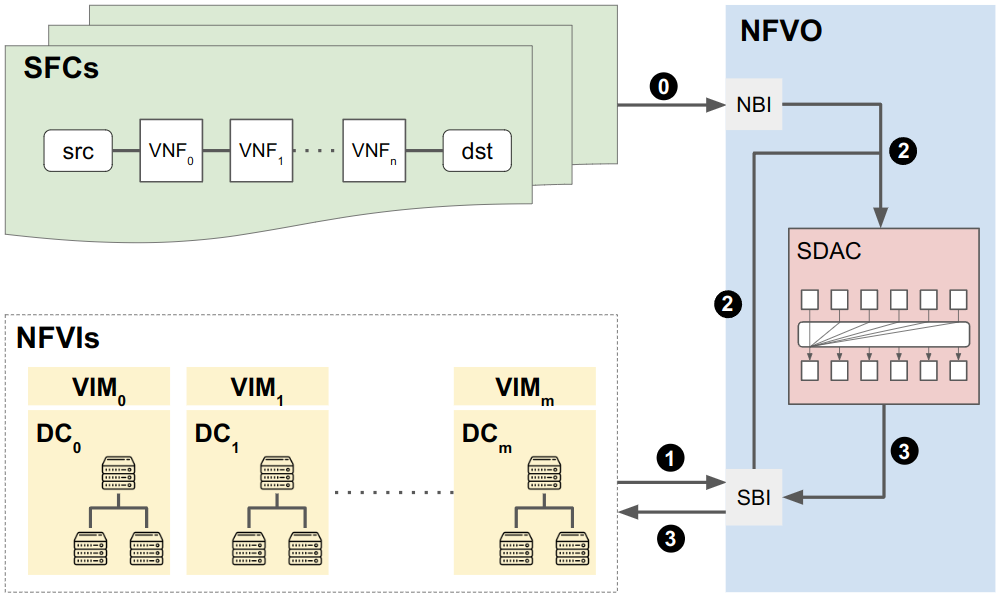}
    \caption{Architecture and flow of the NFVO with the proposed decision-making module (SDAC) for SFC partitioning.}
    \label{fig:MANO}
\end{figure}
We consider that the embedding of partial SFCs within each DC is handled locally by its corresponding Virtual Infrastructure Manager (VIM). The focus of this work lies solely on the partitioning of SFCs at the NFVO level.
The integration of the proposed decision-making component (introduced in Section~\ref{sec:method}) within the NFVO for SFC partitioning is illustrated in Fig.~\ref{fig:MANO}. The process involves the interaction among the SFC requests, the NFVO, and the VIMs. The steps of this process are detailed as follows:

\noindent \textbf{Step 0: Reception of the SFC Request.}  The process starts with the NFVO receiving an SFC request through the Northbound Interface (NBI). The SFC request typically comprises an ordered set of VNFs, denoted as \( \text{VNF}_0, \text{VNF}_1, \ldots, \text{VNF}_n \), which must be deployed between a source (\( \text{src} \)) and a destination (\( \text{dst} \)) target DC. Upon receipt, the NFVO processes the request and extracts corresponding features, which may encompass the resource demands of each VNF, latency constraints, bandwidth requirements, and other QoS parameters.

\noindent\textbf{Step 1: Retrieval of the State of NFVIs.}  In the following, the NFVO retrieves the current state of the Network Function Virtualization Infrastructures (NFVIs) through the Southbound Interface (SBI). The SBI enables communication between the NFVO and the underlying VIMs, denoted as \( \text{VIM}_0, \text{VIM}_1, \ldots, \text{VIM}_m \), each of which manages a corresponding DC (\( \text{DC}_0, \text{DC}_1, \ldots, \text{DC}_m \)). The NFVO gathers monitoring data regarding the available resources across these VIMs, including computational capacity and networking resources of the DCs. 

\noindent\textbf{Step 2: Decision Making using the Joint State.} The NFVO consolidates the processed SFC request (from Step 0) with the current state of the underlying infrastructure (from Step 1) and forwards this joint information to the decision-making module (introduced in Section~\ref{sec:method}).

\noindent\textbf{Step 3: Delivery of Results for Deployment.} Upon receiving the assignments from the decision-making module, the NFVO distributes these results to the respective VIMs through the SBI. The results specify the allocation of VNFs within the SFC across the DCs. For instance, \( \text{VNF}_0 \) might be assigned to \( \text{DC}_1 \), while \( \text{VNF}_1 \) and \( \text{VNF}_2 \) could be co-located to \( \text{DC}_3 \), depending on resource availability and objectives. Each VIM subsequently embeds the assigned VNFs within its corresponding DC, ensuring that the SFC is instantiated and operational in accordance with the partition assignments.
In this multi-domain setting, coordination is handled centrally by the NFVO, which collects aggregated resource statistics (e.g., total CPU and bandwidth) from each domain’s VIM. The SDAC module then jointly processes the full SFC sequence to ensure sequence-aware partitioning decisions across domains. Afterward, the NFVO distributes the assignments to the respective VIMs for local embedding, minimizing the synchronization needs while preserving scalability and privacy.

\begin{table}[ht]
\centering
\caption{Notation Table}
\begin{tabular}{c p{0.7\columnwidth}}
\hline
\toprule
\textbf{Notation} & \textbf{Description} \\ \hline
\midrule
$G_s$ & Substrate network graph with DCs as nodes and inter-DC links as edges. \\
$V_s$ & Set of data centers. \\
$E_s$ & Set of physical links between data centers. \\
$B_l$ & Bandwidth capacity of the physical link $l \in E_s$. \\
$L_l$ & Latency of the physical link $l \in E_s$. \\
$\bar{C}_u$ & Aggregated remaining CPU capacity of DC $u \in V_s$. \\
$\bar{B}_u$ & Aggregated remaining bandwidth capacity of DC $u \in V_s$. \\ 
$G_v$ & SFC request graph with VNFs as nodes and virtual links as edges. \\ 
$V_v$ & Set of VNFs in the SFC. \\ 
$E_v$ & Set of virtual links between VNFs in the SFC. \\ 
$D_v$ & CPU demand of VNF $v \in V_v$. \\ 
$D_e$ & Bandwidth demand of the virtual link $e \in E_v$. \\
$L_{\text{SLA}}$ & E2E latency SLA for an SFC request. \\ 
$x_{v,u}$ & Binary variable: 1 if VNF $v$ is assigned to DC $u$, 0 otherwise. \\ 
$y_{e, l}$ & Binary variable: 1 if the virtual link $e$ is mapped to the physical link $l$, 0 otherwise. \\ 
$\delta$ & Binary variable: 1 if the request is successfully admitted, 0 otherwise. \\ \hline
\bottomrule
\end{tabular}
\label{tab:notation}
\vspace{-1.5em}
\end{table}

\subsection{System Model}
\textbf{Substrate Network.}
The substrate network, denoted as \( G_s = (V_s, E_s) \), is modeled as a graph where 
\( V_s \) is the set of Data Centers (DCs), each \( u \in V_s \) with a CPU capacity of \( C_u \) (in units of available CPU resources) and \( E_s \) is the set of physical links between data centers,  each \( l \in E_s \) with a bandwidth capacity of \( B_l \) and latency \( L_l \). The \textit{inter-DC latency} is modeled as the sum of latencies along the shortest path between two DCs, computed using \textit{Dijkstra’s algorithm}. We consider intra-DC latency as negligible, thus only inter-DC latency contributes to the overall E2E latency.

\textbf{SFC Requests.}
SFC requests are received sequentially over time, with one request arriving during each time slot \( t \). For simplicity, the notation \( t \) is omitted in this section, as we focus on the context within a single time slot. 
Each request is modeled as a 6-tuple $(G_v, t_{\text{arr}}, t_{\Delta}, L_{\text{SLA}}, \textit{src}, \textit{dst})$. \( G_v = (V_v, E_v) \) represents the SFC as a directed graph, with \( V_v \) the set of VNFs,  each \( v \in V_v \) with a CPU demand of \( D_v \) and \( E_v \) the set of virtual links between VNFs,  each \( e \in E_v \) with a bandwidth demand of \( D_e \). \( T_{\text{arr}} \) is the arrival time, \( t_{\Delta} \) is the service lifetime and  \( L_{\text{SLA}} \) is the E2E latency constraint for the request. The \( \textit{src} \) and \( \textit{dst} \) are two auxiliary VNFs with zero resource requirements (\( D_{\text{src}} = D_{\text{dst}} = 0 \)) that represent  the endpoints of the SFC. Accordingly, two additional edges are introduced to the service graph: ($i$) the edge from \( \textit{src} \) to the first VNF \( v_1  \in  V_v \) and ($ii$) the edge from the last VNF \( v_{|V_v|} \in V_v \) to \( \textit{dst} \).


\textbf{Partitioning Decision.}
The \textit{partitioning decision} is made at the NFVO level, which assigns each VNF \( v \in V_v \) to a specific DC \( u \in V_s \):
\begin{equation}\label{eq:decision_var}
x_{v,u} = 
\begin{cases} 
1 & \text{if VNF } v \text{ is assigned to DC } u, \\
0 & \text{otherwise.}
\end{cases}
\end{equation}

At the time of making this decision, the NFVO only has \textit{aggregated information} of the available resources in each DC:
\begin{itemize}
    \item \( \bar{C}_u \): the sum of remaining CPUs in DC \( u \).
    \item \( \bar{B}_u \): the sum of available bandwidth for links associated with DC \( u \).
\end{itemize}

The actual embedding of VNFs inside each DC (i.e., mapping VNFs to individual computing nodes within the DC) is handled by the corresponding VIM of each DC. In contrast, the NFVO is responsible for inter-DC partitioning decisions, determining how VNFs are distributed across different DCs, rather than handling placements within a single DC.
Additionally, the \textit{mapping of virtual links} in \( E_v \) is decided implicitly at the same time when the VNF-to-DC mapping is determined. The shortest paths between assigned DCs are computed using \textit{Dijkstra’s algorithm}, and virtual links are mapped to the corresponding substrate links in \( E_s \), without the need to explicitly decide on link mappings.

\textbf{Resource Constraints.}
\label{subsub:resource constraint}
The following defines the key constraints on CPU and bandwidth capacities, alongside the issue of resource fragmentation.\\
\begin{itemize}
\item \textbf{CPU Capacity.}
The CPU demand of VNFs assigned to a DC must not exceed its aggregated capacity:
\begin{equation}
\sum_{v \in V_v} D_v \cdot x_{v,u} \leq \bar{C}_u, \quad \forall u \in V_s.
\label{eq:constraint_cpu}
\end{equation}

\item \textbf{Bandwidth Capacity.}
There are two types of bandwidth constraints to consider:
\begin{itemize}
    \item \textbf{Intra-DC Bandwidth Capacity Constraint.}  
    The bandwidth demand for virtual links mapped within a DC must not exceed the aggregated bandwidth capacity of that DC:
    \begin{equation}
    \sum_{e \in E_v} D_e \cdot x_{v_1, u} \cdot x_{v_2, u} \leq \bar{B}_u, \quad \forall u \in V_s,
    \label{eq:constraint_bw_intra}
    \end{equation}
    where \( \bar{B}_u \) is the aggregated bandwidth capacity available within DC \( u \), and \( x_{v_1, u} \) and \( x_{v_2, u} \) indicate whether two consecutive VNFs \( (v_1, v_2) \) associated with the virtual link $e$ are assigned to DC \( u \).

    \item \textbf{Inter-DC Bandwidth Capacity Constraint.}
    The bandwidth demand for virtual links mapped between DCs must not exceed the remaining bandwidth capacity of the physical links between them:
    \begin{equation}
    \sum_{e \in E_v} D_e \cdot y_{e, l} \leq B_{l}, \quad \forall l \in E_s,
    \label{eq:constraint_bw_inter}
    \end{equation}
    where \( y_{e, l} \) indicates whether the virtual link \( e \) is mapped to the physical link \( l \).
\end{itemize}
However, meeting the aggregated resource demands is a \textit{necessary but not sufficient condition} for accepting an SFC request due to the possible \textit{resource fragmentation}\footnote{Despite the aggregated resource demands described in formula \eqref{eq:constraint_cpu} and \eqref{eq:constraint_bw_intra} are satisfied for a given request, the request could still be rejected due to resource fragmentation, where the available resources are scattered across multiple computing nodes within a DC, making it impossible to allocate sufficient contiguous resources to meet the request's demands.}.
\end{itemize}

\textbf{E2E Latency Constraint.}
\label{subsub:sla constraint}
The E2E latency $L_{\text{e2e}}$ is computed by traversing the VNFs \( v_1, v_2, \ldots, v_k \) in the SFC sequentially and summing the inter-DC latencies between the DCs they are mapped to:
\begin{equation}
\label{eq:latency}
L_{\text{e2e}} = \sum_{e \in E_v} L_l \cdot y_{e, l}, \quad \forall l \in E_s,
\end{equation}
where \( y_{e, l} = 1 \) if the virtual link \( e \) is mapped to the substrate link \( l \). The E2E latency must satisfy the SLA:
\begin{equation}
L_{\text{e2e}} \leq L_{\text{SLA}}.
\end{equation}

\textbf{Objective.}
\label{subsub:obj}
The goal is to maximize the \textit{long-term acceptance rate} of SFC requests, where acceptance occurs only if all resource and latency requirements are satisfied. This can be formulated as:
\begin{equation}\label{eq:maximize_xy}
\begin{aligned}
& \max_{\{\textbf{x}(t)\}_{t=1}^{T}} \; \frac{1}{T}\sum_{t=1}^{T}\delta_t(\textbf{x}(t)),\\[4pt]
\text{where } 
\delta_t(\textbf{x}(t)) &=
\begin{cases}
1, & \text{if the request at time } t \text{ is accepted},\\
0, & \text{otherwise.}
\end{cases}
\end{aligned}
\end{equation}
Here, \(\textbf{x}(t)\) denotes the assignment of decision variables at time \(t\) defined in~(\ref{eq:decision_var}).
The overall constraints below are the \textit{necessary conditions} for achieving \( \delta_t(\textbf{x}(t)) = 1 \):
\begin{equation}
\begin{aligned}
& \sum_{u \in V_s} x_{v,u} (t) = 1, \quad \forall v \in V_v(t), \\
& \sum_{v \in V_v(t)} D_v (t) \cdot x_{v,u} (t) \leq \bar{C}_u (t), \quad \forall u \in V_s, \\
& \sum_{e \in E_v(t)} D_e(t) \cdot x_{v_1, u}(t) \cdot x_{v_2, u}(t)\\ 
& \quad \leq \bar{B}_u(t), \quad \forall u \in V_s, \\
& \sum_{e \in E_v(t)} D_e (t) \cdot y_{e, l} (t) \leq B_l (t), \quad \forall l \in E_s, \\
& \sum_{e \in E_v(t)} L_l \cdot y_{e, l}(t) \leq L_{\text{SLA}}(t).
\end{aligned}
\end{equation}


\section{Proposed Method}
\label{sec:proposed_method}
Building on the problem formulation, we now present our proposed approach to address the SFCP problem. Given the NP-hard nature of SFCP~\cite{9025750}, we reformulate the problem as a Markov Decision Process (MDP), enabling us to leverage RL techniques for a scalable and efficient solution. This section is organized into two parts: ($i$) the MDP formulation, where we define the states, actions, transitions, and rewards for the problem, and ($ii$) the proposed DRL solution, where we detail our DRL approach.


\subsection{MDP Formulation}
To solve the SFCP problem using reinforcement learning, we first model it as an MDP.
An MDP is defined as a 4-tuple \((\mathcal{S}, \mathcal{A}, \mathcal{T}, \mathcal{R})\), consisting of:

\vspace{1mm}
\noindent \textbf{States (\(\mathcal{S}\)).}
The state captures the characteristics of \textit{individual VNFs} within an SFC request and the \textit{substrate network state}, represented as a vector:
\begin{equation}
s = [s_1, s_2, \ldots, s_n],\quad s_i = \left[ s^{(i)}_{req}, s_{sub} \right],
\label{eq:state}
\end{equation}
where:
\begin{itemize}
    \item \( s_{req} = \left[ D_v, D_e, T_{\text{arr}}, t_{\Delta}, T_{\text{SLA}}, src, dst \right] \) represents the \textbf{request state}, with the individual elements defined in Section~\ref{sec:problem formulation}-B.
    \item \( s_{sub} = \left[ \bar{C}, \bar{B} \right] \) represents the aggregated \textbf{substrate state}, where:
    \begin{itemize}
        \item \( \bar{C}_i \in \left[ \bar{C}_1, \bar{C}_2, \ldots, \bar{C}_m \right] \) is the aggregated remaining CPU resources at data center \( i \).
        \item \( \bar{B}_i \in \left[ \bar{B}_1, \bar{B}_2, \ldots, \bar{B}_m \right] \) is the aggregated remaining bandwidth at data center \( i \).
        \item \( m \) is the total number of data centers.
    \end{itemize}
\end{itemize}
Particularly, since all VNFs in the same SFC share the same substrate state $s_{sub}$, this information is included in each state $s\in S$ to ensure that every token-level action is conditioned on the same network context.
Alternative conditioning strategies are possible, and identifying the most effective approach remains an open design choice beyond the scope of this work.

\vspace{1mm}
\noindent \textbf{Actions (\(\mathcal{A}\)).}
The action corresponds to the joint assignment of all VNFs in the SFC request to target DCs. Formally, let \( V_v = \{v_1, v_2, \dots, v_n\} \) denote the ordered set of VNFs in the SFC, and \( V_s = \{u_1, u_2, \dots, u_m\} \) represent the set of available DCs in the substrate network. The action is defined as:
\begin{equation}
a = [a_1, a_2, \dots, a_n],
\end{equation}
where each \( a_i \) corresponds to the assignment of VNF \( v_i \) and is represented as a one-hot vector of size \( m \):
\begin{equation}
a_i = [a_i^1, a_i^2, \dots, a_i^m] \in \{0, 1\}^m, \quad \sum_{j=1}^m a_i^j = 1.
\end{equation}
Here, \( \arg\max(a_i) = j \) identifies the index \( j \) of the DC \( u_j \) to which the VNF \( v_i \) is assigned. This formulation defines a structured action space over the entire SFC.

\vspace{1mm}
\noindent \textbf{Transitions (\(\mathcal{T}\)).}
The transitions describe how the state evolves as a result of an action. After taking an action \( a \) in state \( s \):
\begin{itemize}
    \item The \textbf{substrate state} \( s_{sub} \) is updated to reflect the resources consumed by the embedded VNFs and their virtual links. If the embedding fails, the substrate state remains unchanged. Additionally, resources consumed by \textit{expired services} are released back into the substrate.
    \item The \textbf{request state} \( s_{req} \) transitions to the next incoming SFC request, with its resource demands, arrival time, service lifetime, and SLA.
\end{itemize}
The detailed steps of this process are provided in Section~\ref{sec:eval}.

\vspace{1mm}
\noindent \textbf{Rewards (\(\mathcal{R}\)).}  
The reward evaluates the outcome of the partitioning decision:
\begin{equation}
r(s, a) = 
\begin{cases} 
1, & \text{if the request is accepted}, \\
0, & \text{otherwise}.
\end{cases}
\label{eq:reward}
\end{equation}
A reward of \(1\) is assigned if the SFC request is accepted, meaning all VNFs are successfully embedded, and the E2E latency satisfies the SLA. A reward of \(0\) is assigned if the request is rejected, either due to a failure in embedding any VNF or a violation of the SLA.
Note that this immediate reward reflects the acceptance indicator \(\delta\) in~\eqref{eq:maximize_xy}.



\subsection{The Proposed Framework: SDAC}\label{sec:method}
\begin{figure*}[h]
     \centering
     \includegraphics[width=2\columnwidth]{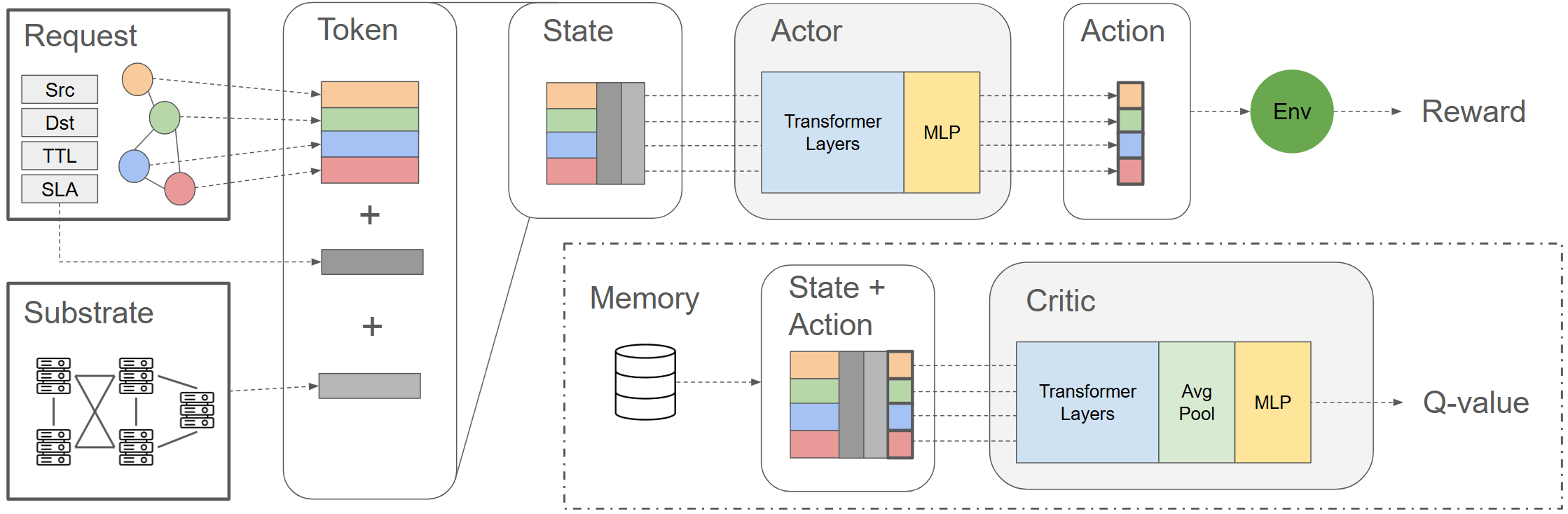}
    \caption{Overview of the SDAC framework with Transformer-based architecture for sequence-aware SFC partitioning.}
    \label{fig:drl}
    \vspace{-3mm}
\end{figure*}

We propose a DRL framework, named \textbf{S}equence-aware \textbf{D}ifferentiable \textbf{A}ctor-\textbf{C}ritic
RL, for efficient SFC partitioning.
Our approach incorporates Transformer encoders in both the actor and critic networks, leveraging their ability to model inter-dependencies among VNFs and make informed, coordinated decisions, as shown in Fig.~\ref{fig:drl}. Below, we detail the components of the framework.

\begin{algorithm}
\caption{Learning Framework of SDAC}
\label{alg:drl_sfcp}
Initialize actor network \( \pi_\theta \) and critic network \( Q_\phi \)

Initialize target networks \( \pi_{\theta'} \gets \pi_\theta \) and \( Q_{\phi'} \gets Q_\phi \)

Initialize replay buffer \( \mathcal{D} \)

Initialize perturbation factor $\epsilon$

\For{\( \text{episode} = 1 \) \KwTo \( N \)}{
    Reset environment and observe the initial state \( s_0 \)
    
    \For{\( t = 1 \) \KwTo \text{total number of SFC requests}}{
        \textbf{// Parallel Action Selection}
        \[
        a_t \gets  \epsilon \text{-LoPe}(s_t, \pi_\theta, \epsilon)
        \]
        
        \textbf{// Environment Interaction}
        
        Run action \( a_t \), get reward \( r_t \) and next state \( s_{t+1} \)
        
        \textbf{// Transition Storing}
        
        Store transition \( (s_t, a_t, r_t, s_{t+1}) \) in \( \mathcal{D} \)
        
        \textbf{// Critic Update}
        
        Sample \( M \) mini-batch transitions \( (s, a, r, s') \)
        
        from \( \mathcal{D} \) and compute target Q-values:
        \[
        y \gets r + \gamma Q_{\phi'}(s', \pi_{\theta'}(s'))
        \]
        
        Update critic by minimizing the loss:
        \[
        \mathcal{L}_{\text{critic}} = \frac{1}{M} \sum \left( Q_\phi(s, a) - y \right)^2
        \]
        
        \textbf{// Actor Update}
        
        Update actor by minimizing the loss:
        \[
        \mathcal{L}_{\text{actor}} = -\frac{1}{M} \sum Q_\phi(s, \pi_\theta(s))
        \]\\
        
        \textbf{// Target Network Update}
        
        Update target networks:
        \[
        \phi' \gets \tau \phi + (1 - \tau) \phi'\]
        \[ \theta' \gets \tau \theta + (1 - \tau) \theta'\]
        
    }
}
\Return \( \pi_\theta \)
\end{algorithm}

\textbf{Transformer-Based Actor-Critic Framework.}
The Transformer architecture~\cite{vaswani2017attention}, originally proposed for NLP tasks, is known for its ability to model relationships between input tokens using self-attention.
A Transformer encoder layer comprises two main components: a multi-head self-attention mechanism and a Multi-Layer Perceptron (MLP).
In the SFCP problem, VNF placement decisions are mutually coupled: assigning one VNF to a data center affects the feasibility of placing others through shared resource capacities, bandwidth routing, and E2E latency constraints. As a result, request acceptance depends on the joint assignment of all VNFs rather than on independent local decisions. The policy must therefore evaluate each VNF conditioned on the other VNFs within the same SFC.
The self-attention mechanism enables this joint conditioning by allowing each VNF representation to incorporate information from all other VNFs simultaneously, producing coordinated partitioning decisions. We represent each VNF as an input “token” and process the SFC as an ordered sequence using a Transformer encoder. Note that the token representation serves only as an input structuring mechanism and does not imply linguistic modeling.
Unlike in autoregressive linguistic tasks, no causal masking is applied.
\begin{itemize}
    \item \textbf{Self-Attention for Inter-Dependency Modeling.}
Self-attention enables the model to capture inter-dependencies among VNFs by allowing each VNF to attend to all others in the sequence. Given a sequence of $n$ VNF tokens
\begin{equation}
S = [s_1, s_2, \ldots, s_n] \in \mathbb{R}^{n \times d}, \label{eq:states}
\end{equation}
self-attention produces an updated representation $S'$ by computing context-aware combinations of tokens.
The mechanism first projects the input into query, key, and value spaces:
\begin{equation}
Q = S W_Q, \quad K = S W_K, \quad V = S W_V,
\end{equation}
where $W_Q, W_K, W_V$ are learnable parameters. Attention weights are then computed as
\begin{equation}
E = \text{softmax}\left( \frac{QK^\top}{\sqrt{d_k}} \right),
\end{equation}
and applied to the values to obtain the output:
\begin{equation}
S' = E V.
\end{equation}

By aggregating information across all VNFs, self-attention captures global dependencies within the SFC, enabling partitioning decisions for each VNF to account for the resource demands of the entire sequence.
    \item \textbf{Actor Network with Transformers.}  
    The actor uses Transformer encoder layers to process the sequence of VNFs in the SFC.
    The input to the actor is the set of tokens of the entire SFC $S$, as defined in~(\ref{eq:states}). This matrix is passed through the Transformer encoder layers. The actor outputs a probability distribution over the target DCs for each VNF, representing the policy \( \pi_\theta(s) \). Specifically, the policy can be expressed as:
    \begin{equation}\label{eq:actions}
    \pi_\theta(S) = [\pi_\theta(s_1), \pi_\theta(s_2), \dots, \pi_\theta(s_n)],
    \end{equation}
    where \( \pi_\theta(s_i)=a_i \) is a one-hot vector that encodes the selected action over the available DCs for the $i$-th VNF.
    \item \textbf{Critic Network with Transformers.}  
    The critic  \( Q_\phi(s, a) \) also adopts Transformer encoder layers to process the sequence of VNFs and their corresponding partitioning decisions. Instead of evaluating each decision individually, the critic evaluates the entire sequence of state-actions for the SFC as a whole. In particular, the input to the critic encodes the entire SFC, represented by the state matrix \( S \) defined in~(\ref{eq:states}), concatenated with the partitioning decisions \( A \), where \( A = [a_1, a_2, \dots, a_n] \) represents the assignment of each VNF to a specific DC.
    Formally, the input to the critic is defined as:
    \begin{equation}\label{eq:concate_h}
    H = [h_1, h_2, \dots, h_n],
    \end{equation}
    where \( h_i = [s_i; a_i] \), \( s_i \) is the feature of the \( i \)-th VNF, \( a_i \) is the one-hot representation of the partitioning decision for \( i \)-th VNF, and \( [\cdot; \cdot] \) denotes concatenation.
    The output of the Transformer is then pooled into a single fixed-length vector using element-wise mean pooling. This approach borrows the idea from Universal Sentence Encoder (USE) architecture~\cite{cer2018universal} commonly used in NLP tasks, where variable-length sequences are mapped into fixed-length representations.
    The pooled representation is then passed through a linear layer to compute the Q-value, providing a holistic evaluation of the decisions for the entire SFC.
    
    
    \item \textbf{Target Networks}  
    To stabilize training, two target networks are employed for both the actor \( \pi_{\theta'}(s) \) and critic \( Q_{\phi'}(s, a) \). A target network is a separate network used to compute the target values for updates to the main network. This separation prevents the network from chasing its own moving predictions, which can lead to instability or divergence in training. The target network's parameters are updated less frequently or with a smoothing mechanism (e.g., Polyak averaging) to improve convergence:
    \begin{equation}
    \phi' \gets \tau \phi + (1 - \tau) \phi', \quad \theta' \gets \tau \theta + (1 - \tau) \theta',
    \end{equation}
    where \( 0 < \tau \ll 1 \) is the target update rate.
\end{itemize}


While several studies have proposed sophisticated inter-decision communication mechanisms to improve the coherence of VNF placements, these approaches often introduce significant complexity, complicating both optimization and implementation. In contrast, the use of Transformers offers an elegant and well-validated solution for achieving coordination without unnecessary complexity.

\textbf{Training Process and Objectives.}  
The training process adopts the standard actor-critic framework, which uses separate loss functions for the actor and critic networks. The complete training steps are given in Alg.~\ref{alg:drl_sfcp} and~\ref{alg:e_blending}.

\begin{itemize}
    \item \textbf{Critic Loss.}  
    The critic is trained to minimize the Mean Squared Error (MSE) between the predicted Q-value \( Q_\phi(s, a) \) and the target Q-value \( y \). The target Q-value is computed using the Bellman equation:
    \begin{equation}
    y = r + \gamma Q_{\phi'}(s', \pi_{\theta'}(s')),
    \end{equation}
    where \( r \) is the immediate reward, \( \gamma \) is the discount factor, and \( Q_{\phi'} \) and \( \pi_{\theta'} \) are the target critic and actor networks, respectively.
    The critic loss is defined as:
    \begin{equation}
    \mathcal{L}_{\text{critic}} = \mathbb{E} \left[ \left( Q_\phi(s, a) - y \right)^2 \right].
    \end{equation}
    
    \item \textbf{Actor Loss.}  
    The actor is trained to maximize the Q-value of the actions it generates, as evaluated by the critic. This is equivalent to minimizing the negative expected Q-value:
    \begin{equation}
    \mathcal{L}_{\text{actor}} = -\mathbb{E} \left[ Q_\phi(s, \pi_\theta(s)) \right].
    \end{equation}
    By leveraging the critic's evaluation, the loss function guides the actor to update its policy in the direction that increases the Q-value, ultimately improving decision-making performance.
    
    
    

\end{itemize}

\textbf{Continuous Action Representation.}
\label{actionselection}
In the SFCP problem, the action space is inherently discrete, since each VNF must be assigned to exactly one domain or DC. A go-to approach would be to treat the action as categorical and rely on sampling to select a target domain. However, such sampling-based approaches introduce \textit{non-differentiability}, which prevents E2E gradient-based training of the actor-critic framework. This step is essential because the USE-style critic evaluates the entire sequence holistically, and differentiable actions are required for providing feedback to guide the actor.
To address this, we adopt a continuous relaxation of the discrete action space by applying a softmax function to the actor’s logits. This produces a “relaxed one-hot” vector that approximates the discrete decision while remaining differentiable~\cite{jang2017categorical}, allowing gradients to flow directly through the actor-critic framework. 
The relaxation also improves coordination across multiple VNFs, since joint assignments can be optimized coherently rather than attributing credit to each decision independently. At inference, we apply the argmax operation to the softmax probabilities, yielding discrete assignments that are fully consistent with the original problem formulation.
Thus, the continuous representation provides a principled bridge between the discrete nature of SFC partitioning and the requirements of scalable DRL, preserving validity while enabling efficient optimization and simple training processes.

\textbf{Epsilon-Logit Perturbation ($\bm{\epsilon}$-LoPe).}
Since discrete actions are encoded as continuous representations, the traditional $\epsilon$-greedy exploration strategy becomes inapplicable. To this end, we introduce $\epsilon$-LoPe, a noise injection strategy that facilitates exploration.
Instead of selecting a random action with probability \( \epsilon \), $\epsilon$-LoPe perturbs the logits before applying softmax, imposing the exploration at the representation level rather than in the final actions. 
Specifically, at each decision step, a Gaussian noise \( \eta \sim \mathcal{N}(0, 1) \) is scaled by the decaying perturbation factor \( \epsilon \in [0, 1] \) and a constant $c$, and then added to the logits \( z_t \):
\begin{equation}\label{eq:blending}
    \tilde{z}_t = z_t + \epsilon \cdot c \cdot \eta,
\end{equation}
where \( z_t \) represents the pre-softmax logits generated by the actor network. The perturbed logits \( \tilde{z}_t \) are then passed through a softmax function to produce the final action:
\begin{equation}
    a_t = \text{softmax}(\tilde{z}_t).
\end{equation}
The perturbation factor \( \epsilon \) decreases over episodes, encouraging high exploration in early training while gradually moving toward deterministic policy as training progresses.

To regulate the impact of noise injection, we apply z-score normalization to logits $z_t$ across features per sample before adding the scaled noise, as it regulates the scale of logits to remain comparable to the noise $\eta$ sampled from the standard normal distribution.
In addition, the noise $\eta$ is scaled by a constant $c$, which serves as a hyperparameter to control the magnitude of the noise.
The step in~(\ref{eq:blending}) is thus updated to:
\begin{equation}\label{eq:norm_add}
    \tilde{z}_t = \frac{z_t-\mu(z_t)}{\sigma(z_t)} + \epsilon \cdot c \cdot  \eta,
\end{equation}
where $\mu(z_t)$ and $\sigma(z_t)$ are the mean and standard deviation calculated across the elements of the logit vector $z_t$.
The complete process is given in Alg.~\ref{alg:e_blending}.
By introducing a tunable perturbation at the logit level rather than overriding the final action, $\epsilon$-LoPe enables controllable action exploration while preserving differentiability.
An ablation study that examines the effectiveness of the proposed $\epsilon$-LoPe exploration strategy is given in Section~\ref{ablation}.
\begin{algorithm}[h]
\caption{$\epsilon$-LoPe}
\label{alg:e_blending}
\KwIn{state \( s_t \), actor \( \pi_\theta \), perturbation factor \( \epsilon \)}

\textbf{// Compute Pre-Softmax Logits}

Obtain raw logits from $\pi_\theta$ before applying softmax:
\[
z_t \gets \pi_\theta^{\text{logit}}(s_t)
\]



\textbf{// Perturb Logits with Gaussian Noise}

Add scaled noise to normalized logits:
\[
\tilde{z}_t \gets \frac{z_t-\mu(z_t)}{\sigma(z_t)} + \epsilon \cdot c \cdot  \eta ,\quad \eta \sim \mathcal{N}(0, 1)
\]

\textbf{// Compute the Final Action}

Apply softmax to obtain the final action:
\[
a_t \gets \text{softmax}(\tilde{z}_t)
\]

\Return \( a_t \)
\end{algorithm}

\textbf{Asymptotic Return Normalization.} 
Training Transformer-based DRL in large-scale environments often suffers from instability due to extreme or highly variable reward magnitudes. Existing techniques such as reward clipping~\cite{mnih2013playing} or z-score normalization can alleviate this but might introduce task-dependent heuristics or inaccurate gradient signals. To address this in a lightweight and principled way, we propose Asymptotic Return Normalization (ARN). The key idea is to normalize immediate rewards such that the maximum possible discounted return remains bounded in \([0,1]\).
We derive the normalizing factor as follows: Given an immediate reward \( r_t \in [0,1]\), the expected cumulative discounted reward \( G_t \) over \( N \) steps with discount factor $\gamma$ is defined as a finite geometric series with a common ratio of \( \gamma \), where \( 0 < \gamma < 1 \). The sum of this series is given by:
\begin{equation}
G_t = \sum_{k=0}^{N-1} \gamma^k r_{t+k}.
\end{equation}
Since the rewards are bounded as \( 0 \leq r_t \leq 1 \), the maximum possible value of \( G_t \) occurs when all rewards take their upper bound (i.e., \( r_t = 1 \) for all steps):
\begin{equation}
G_{\max} = \sum_{k=0}^{N-1} \gamma^k,
\end{equation}
and using the formula for the sum of a finite geometric series:
\begin{equation}
G_{\max} = \frac{1 - \gamma^N}{1 - \gamma}.
\end{equation}
To determine the normalization factor, we calculate the maximum possible return as \( N \to \infty \). Since \( 0 < \gamma < 1 \), the term \( \gamma^N \) approaches 0 as \( N \to \infty \), giving us the infinite sum:
\begin{equation}
G_{\max} = \sum_{k=0}^{\infty} \gamma^k = \frac{1}{1 - \gamma}.
\end{equation}
Thus, the return should ideally be normalized as:
\begin{equation}
G_t' = \frac{G_t}{G_{\max}} =  (1 - \gamma) G_t = (1 - \gamma) \sum_{k=0}^{N-1} \gamma^k r_{t+k}.
\end{equation}
However, since we employ temporal difference learning approach, we do not explicitly compute the full return \( G_t \) at each step. Instead, we apply the normalization directly to the immediate reward to achieve the same effect. That is, we define the normalized reward as:

\begin{equation}
r_t' = (1 - \gamma) r_t.
\end{equation}
This keeps the return within \([0,1]\), preventing gradient explosion while preserving the reward structure, and removes the need for ad-hoc techniques such as warm-up or pretraining~\cite{brown2020language, steiner2022how}. ARN is task-agnostic and applicable to any DRL environment with a known maximum return.
An ablation study evaluating the impact of the proposed ARN, along with its interaction with other design components, is presented in Section~\ref{ablation}.



\section{Experimental Setup} 
\label{sec:perf_eval}
To evaluate the effectiveness of the proposed framework, we conduct a comprehensive performance assessment. This section is structured into three parts: ($i$) the evaluation scenario, which provides details about the simulation environment; ($ii$) the comparison algorithms, including SoTA (meta-)heuristic and DRL-based approaches; ($iii$) the configurations of hyper-parameters for the algorithms.

\subsection{Evaluation Scenario}
\label{sec:eval}
To capture the inherent heterogeneity and variability of multi-domain mobile network environments, we adopt a generalized simulation approach rather than modeling a specific deployment scenario. Domain characteristics such as resource capacities and service lifetimes varied over a wide range of values. Multiple independent simulation runs are conducted to ensure that the diversity and dynamics of domain conditions are sufficiently reflected.
The simulation environment consists of a total of $10,000$ linear SFC requests, each comprising $2$ to $10$ VNFs, uniformly distributed~\cite{6226390}. Each VNF requires between $5$\% and $20$\% of the CPU resources, also uniformly distributed~\cite{fake10107385}. The arrival of SFC requests follows a Poisson process, with an average rate of $\lambda = 0.05$ requests per time unit.
Although the evaluation in this research considers linear service chains for comparability with prior works~\cite{fake10107385, 9685162-SCHEMA}, our method can be directly extended to more complex topologies\footnote{The SDAC framework is not restricted to linear chains. The self-attention mechanism does not assume a linear structure; it can learn dependencies among VNFs regardless of whether the service graph is a chain or a DAG. This makes the approach naturally applicable to more general service topologies.}.
To model evolving traffic intensity over time, the average rate $\lambda$ is modulated by a sinusoidal function as in~\cite{hsu2025rails}.
The service lifetime for each request is sampled from an exponential distribution with a mean of $1,000$ time units~\cite{6226390}. The SLA for each request, expressed in terms of E2E latency requirements, is sampled uniformly from the range \( [2, 4] \) time unit. 
The auxiliary VNFs, $src$ and $dst$, which represent the endpoints of the SFC, are independently assigned to randomly selected DCs.
The substrate network comprises $5$ data centers, modeled as a random graph initialized using the \textit{Erdős–Rényi model}, where each possible edge occurs with a probability of $0.5$. Each DC contains a number of computing nodes sampled uniformly from the set $\{32, 64, 128, 256\}$.
The initial load on these computing nodes varies between $70$\% and $100$\%, also sampled uniformly. The bandwidth capacity of all links is assumed to be sufficient at all times~\cite{fake10107385}. 
For determining the E2E latency of a service request, \textit{Dijkstra's algorithm} is applied to compute the shortest path between the DCs hosting consecutive VNFs once all VNFs in an SFC are assigned. 
Each inter-DC link is assigned a latency of $1$ time unit, i.e., $L_l=1$ in formula~(\ref{eq:latency}), while the latency for intra-DC links is assumed to be negligible. The complete simulation operates as follows:

\begin{enumerate}
    \item \textbf{Process Incoming Request.}  
    The system receives the next SFC request along with its partitioning assignment, as predicted by the algorithm. This assignment specifies the target DCs for embedding the VNFs within the SFC.

    \item \textbf{Handle Expired Services.}  
    The system checks for any expired services, releases their allocated resources, and updates the substrate network state to reflect the latest available resources.

    \item \textbf{Estimate Resource Feasibility.}  
    The system evaluates whether the resources required by the SFC, based on the assignment from Step 1 and the formulas in Section~\ref{subsub:resource constraint}, are available in the target DCs. If the resource requirements cannot be met, the request is rejected, and the system proceeds to Step 1 to handle the next request.

    \item \textbf{Check SLA Compliance.}
    If the resource requirements are satisfied in Step 3, the system verifies whether the E2E latency of the SFC meets the specified SLA using formulas in Section~\ref{subsub:sla constraint}. If the latency requirement is fulfilled, the request is accepted, and the target DCs proceed with embedding the corresponding VNFs. Othrwise, the request is rejected, and the system moves to Step 1 for the next request.
\end{enumerate}
This flow ensures that each incoming request is processed in a structured manner, accounting for resource availability and SLA compliance, while dynamically managing the system's state to handle evolving service demands.


\subsection{Comparison Solutions}
To evaluate the effectiveness of our proposed solution, we compare it with five representative algorithms, including both heuristic and SoTA methods. These algorithms are described as follows:
\begin{itemize}
    \item \textbf{Greedy Policy (GP).}
    In this greedy heuristic-based approach, used as a baseline for comparison, the SFC is not partitioned. Instead, the entire SFC is assigned to a single DC that has the maximum remaining resource capacity at the time of decision-making. While this method reduces the complexity of partitioning, it may lead to inefficient utilization of resources, particularly for longer SFCs, and can result in higher rejection rates due to capacity constraints. This approach serves as a baseline for evaluating more advanced partitioning methods.
    \item \textbf{Iterated Local Search (ILS).}
    ILS is a meta-heuristic optimization method that iteratively refines solutions by applying local search techniques with perturbation to escape local optima~\cite{Lourenço2003}. This SoTA solution is well-suited for partitioning and embedding problems, as demonstrated in prior research~\cite{6226390}.
    The original ILS algorithm, however, considers only resource constraints in the optimization process, without accounting for latency-related metrics. To address this limitation, we modify the algorithm to prioritize minimizing E2E latency for the current SFC request while adhering to resource constraints. The modified ILS framework operates as follows:
    \begin{enumerate}
    \item \textbf{Initial Assignment.}  
    Generate an initial assignment for the SFC by randomly mapping each VNF to an available DC.
    
    \item \textbf{Perturbation.}  
    With a given probability, each VNF's assigned DC is altered to a randomly selected DC.

    \item \textbf{Local Search.}  
    Perform local search by selecting a random VNF from the SFC and iterating over all available DCs. The VNF is temporarily reassigned to each DC, and the DC that yields the minimum latency while satisfying resource constraints in~(\ref{eq:constraint_cpu}) is chosen for reassignment.
    
    \item \textbf{Acceptance Criterion.}  
    Accept the newly obtained assignment if it achieves a lower E2E latency compared to the current best assignment. Otherwise, retain the existing best assignment.
    
    \item \textbf{Iterative Refinement.}  
    Repeat steps $2–4$ until the E2E latency constraint is met, or the predefined maximum number of iterations is reached.
\end{enumerate}
    Despite its strengths, the nature of ILS does not account for long-term performance, which may lead to over-provisioning of resources for the individual request. This can result in future requests being rejected due to insufficient resources, thereby impacting the overall system efficiency over time.
    It serves as a strong optimization-based benchmark for comparison.

    \item \textbf{Risk-Aware Iterated Local Search (RAILS).}
    RAILS extends the standard ILS by incorporating online risk models to predict the acceptance probability of assignments for each domain~\cite{hsu2025rails}. Unlike ILS, which evaluates resource feasibility based on aggregated available resources, RAILS maintains a DC-specific risk model that dynamically estimates the likelihood of acceptance given ($i$) the aggregated available resources, ($ii$) total resource demands, and ($iii$) the total number of VNFs to be placed there. 
    Specifically, for DC $d$, let \( P \) denote the probability of acceptance for assigning a set of VNFs, given the input features \( \mathbf{x} \) (the DC subscript $d$ is omitted as we here focus on a single DC).
    The risk model can be expressed as:
    \begin{equation}
        P(\mathbf{x}) = f_{\theta}(R, D_{\text{total}}, N_{\text{total}}),
    \end{equation}
    where the feature vector \( \mathbf{x} = (R, D_{\text{total}}, N_{\text{total}}) \) consists of the aggregated available resources at DC \( d \), denoted as \( R \); the total resource demands of the VNFs to be placed at DC \( d \), denoted as \( D_{\text{total}} \); and the total number of VNFs to be assigned to DC \( d \), denoted as \( N_{\text{total}} \). The function \( f_{\theta} \) is a NN parameterized by \( \theta \).
    The risk model is built upon historical feedback, allowing the model to learn and adapt over time. While aggregated resource constraints must be met for a mapping to succeed, they alone do not guarantee acceptance. Risk models help capture these subtler admission control dynamics, which would otherwise be inaccurately estimated if relying solely on aggregate metrics.
    The workflow of RAILS closely resembles ILS, with the key difference being that feasibility evaluation is handled by risk models rather than directly assessing aggregated available resources as defined in~(\ref{eq:constraint_cpu}). These risk models are updated periodically after a certain number of steps to reflect changing system conditions.
    The training of the risk model \( f_{\theta} \) is performed by minimizing the cross-entropy loss, using historical feedback as labels. Formally, for a given DC \( d \), let \( y_i \in \{0, 1\} \) denote the binary feedback (1 for acceptance, 0 for rejection) for the sample $i$, and let \( \hat{y}_i = f_{\theta}(\mathbf{x}_i) \in [0, 1] \) be the predicted probability given feature vector \( \mathbf{x}_i \). The loss function is defined as:
    \begin{equation}
        -\frac{1}{K} \sum_{i=1}^K \left[ y_i \log(\hat{y}_i) + (1 - y_i) \log(1 - \hat{y}_i) \right],
    \end{equation}
    where \( K \) is the number of feedback samples. The parameters \( \theta \) are updated by minimizing this loss.
    The up-to-date risk models are then used to predict the acceptance probability of a request at a given DC. 
    If the predicted probability $\hat{y}$ exceeds a predefined threshold $\rho_{\text{risk}}$, the assignment is considered feasible.
    This approach bridges the gap between meta-heuristic-based optimization and data-driven decision-making, making it a more effective solution for SFC partitioning.
    Despite its advantages, RAILS, like ILS, does not explicitly optimize for long-term performance. While it enhances decision-making in the short term, it does not proactively manage resource availability across future requests.
    
    \item \textbf{Sequential DDQN (seqDDQN).}
    This SoTA DRL-based solution employs Double Deep Q-Network (DDQN) to sequentially map VNFs one by one to target data centers (see Alg.~\ref{alg:sddqn}).
    As one of the leading DRL methods for tackling the SFCP problem, seqDDQN inherently optimizes for long-term performance via Q-learning.
    The decision for each VNF is made with access to the action of the preceding VNF, allowing for a degree of coordination between assignments.
    In particular, the next state $s'$ in the transition represents the entire SFC of the subsequent request. The Q-value of the next state-action is estimated by averaging the corresponding maximum Q-values across all VNFs in the next state.
    However, its sequential nature can result in longer decision-making times and may miss globally optimal solutions due to its stepwise approach.
    In this work, reward signals are shared between decisions within the same SFC. This method serves as a modern DRL-based benchmark.
 
    \item \textbf{Parallel DDQN (paraDDQN).}
    This approach also utilizes a DDQN-based policy as in seqDDQN, but maps all VNFs in parallel (see Alg.~\ref{alg:pddqn}).
    By processing all VNFs simultaneously, paraDDQN achieves significant improvements in decision-making speed. However, the placements are determined independently, without explicit coordination among VNFs, which limits its ability to account for inter-dependencies effectively.
    To address this challenge, several studies have introduced sophisticated inter-decision communication mechanisms to enhance coordination among VNF assignments. Although these methods improve mapping coherence, they also add considerable complexity to the optimization process, increasing the difficulty of implementation and management.
    In this work, we adopt shared reward across agents, which enables agents to learn implicit coordination during training~\cite{matignon2012independent}.
    The key difference from seqDDQN is that predictions can be made in parallel, as the state of each VNF does not depend on the action of the preceding VNFs. This method achieves a balance between performance and computational efficiency, making it a robust modern DRL-based benchmark for comparison.
    \end{itemize}
We compare our proposed method with the algorithms presented above, which include both (meta-)heuristic and DRL-based approaches. 
Each approach has unique strengths and design choices. Their key properties and distinguishing features are summarized in Table~\ref{tab:algo-properties}.
    

\begin{algorithm}[h]
\caption{Sequential DDQN}
\label{alg:sddqn}

Initialize Q-network $Q_\phi$

Initialize target network $Q_{\phi'} \gets Q_\phi$

Initialize replay buffer $\mathcal{D}$

\For{\( \text{episode} = 1 \) \KwTo \( N \)}{
    Reset environment and observe the initial state $s_0$
    
    \For{$t = 1$ \KwTo \text{total number of SFC requests}}{
            \textbf{// Sequential Action Selection}
            
            $a_t \gets \{\}$
            
            $a_{\text{prev}} \gets \text{None}$
            
            \For{each VNF $v$ in the SFC}{
                $s_v \gets [s_v; a_{\text{prev}}]$
                
                $a_v \gets \arg\max_{a} Q_{\phi}(s_v, a)$ with $\epsilon$-greedy
                
                $a_t \gets a_t \cup \{a_v\}$
                
                $a_{\text{prev}} \gets a_v$
                }
            \textbf{// Environment Interaction}
            
            Run action $a_t$, get reward $r_t$ and next state $s_{t+1}$
            
            \textbf{// Transition Storing}
            
                \For{each VNF $v$ in $s_t$}{
                    Store transition $(s_v, a_v, r_t, s_{t+1})$ in $\mathcal{D}$
                }
            \textbf{// Q-network Update}
            
            Sample $M$ mini-batch transitions $(s, a, r, s')$
            
            from $\mathcal{D}$ and compute target Q-value:
            
            $q \gets \{\}$\\
            $a_{\text{prev}} \gets \text{None}$\\
                \For{each VNF $v$ in $s'$}{
                    $s_v \gets [s_v; a_{\text{prev}}]$\\
                    $a_v \gets \arg\max_{a} Q_{\phi}(s_v, a)$\\
                    $q_v \gets Q_{\phi'}(s_v, a_v)$\\
                    $q \gets q \cup \{q_v\}$\\
                    $a_{\text{prev}} \gets a_v$
                }
            \[y \gets r + \gamma \cdot \text{mean}(q)\]\\
            Update critic by minimizing the loss:
            \[\mathcal{L}_{\text{critic}} = \frac{1}{M} \sum (Q_\phi(s, a) - y)^2\]\\
            \textbf{// Target Network Update}\\
            Update target networks:
            \[\phi' \gets \tau \phi + (1 - \tau) \phi'\]
    }
}
\Return \( Q_{\phi} \)
\end{algorithm}

\begin{algorithm}[h]
\caption{Parallel DDQN}
\label{alg:pddqn}

Initialize Q-network $Q_\phi$\\
Initialize target network $Q_{\phi'} \gets Q_\phi$\\
Initialize replay buffer $\mathcal{D}$

\For{\( \text{episode} = 1 \) \KwTo \( N \)}{
    Reset environment and observe the initial state $s_0$\\
    \For{$t = 1$ \KwTo \text{total number of SFC requests}}{
            \textbf{// Parallel Action Selection}\\
            $a_t \gets \arg\max_{a} Q_{\phi}(s_t, a)$ with $\epsilon$-greedy\\
            \textbf{// Environment Interaction}\\
            Run action $a_t$, get reward $r_t$ and next state $s_{t+1}$\\
            \textbf{// Transition Storing}\\
                \For{each VNF $v$ in $s_t$}{
                    Store transition $(s_v, a_v, r_t, s_{t+1})$ in $\mathcal{D}$
                }
            \textbf{// Q-network Update}\\
            Sample $M$ mini-batch transitions $(s, a, r, s')$\\ from $\mathcal{D}$
            and compute target Q-value:\\
            $q \gets \{\}$\\
                \For{each VNF $v$ in $s'$}{
                    $a_v \gets \arg\max_{a} Q_{\phi}(s_v, a)$\\
                    $q_v \gets Q_{\phi'}(s_v, a_v)$\\
                    $q \gets q \cup \{q_v\}$
                }
            \[y \gets r + \gamma \cdot \text{mean}(q)\]
            
            Update critic by minimizing the loss:
            \[\mathcal{L}_{\text{critic}} = \frac{1}{M} \sum (Q_\phi(s, a) - y)^2\]
            
            \textbf{// Target Network Update}
            
            Update target networks:
            \[\phi' \gets \tau \phi + (1 - \tau) \phi'\]
    }
}
\Return \( Q_{\phi} \)
\end{algorithm}

\begin{table*}[t]                                
  \caption{Qualitative properties of the comparison algorithms.}
  \label{tab:algo-properties}
  \centering
  \renewcommand{\arraystretch}{1.05}
  \setlength\tabcolsep{5pt}                      
  \begin{tabular}{|l|c|c|c|c|c|c|c|}
    \hline
    \textbf{Algorithm} & \textbf{Opt-based} & \textbf{ML-based} & \textbf{Long-term Performance} & \textbf{Parallel Decision} & \textbf{Sequence Aware} & \textbf{(Re-)Training Required} & \textbf{Risk Model} \\
    \hline
    GP          & \ding{51} &   &   & \ding{51} & \ding{51} &   &   \\
    \hline
    ILS         & \ding{51} &   &   & \ding{51} & \ding{51} &   &   \\
    \hline
    RAILS       & \ding{51} & \ding{51} &   & \ding{51} & \ding{51} & \ding{51} & \ding{51} \\
    \hline
    seqDDQN     &   & \ding{51} & \ding{51} &   & \ding{51} & \ding{51} &   \\
    \hline
    paraDDQN    &   & \ding{51} & \ding{51} & \ding{51} &   & \ding{51} &   \\
    \hline
    SDAC    &   & \ding{51} & \ding{51} & \ding{51} & \ding{51} & \ding{51} &   \\
    \hline
  \end{tabular}
  \vspace{-1ex}
  \begin{flushleft}
    \footnotesize
    \textbf{Legend.} Opt-based~= classical optimization / (meta‑)heuristic;
    ML-based = uses machine learning;
    Long-term Performance = explicit long‑term reward optimization; 
    Parallel Decision = decisions for all VNFs in one step;
    Sequence Aware = decision for one VNF explicitly accounts for others in the input sequence;
    (Re-)Training Required = requires online or offline (re‑)training;  
    Risk Model = maintains explicit admission‑risk models.
  \end{flushleft}
  \vspace{-3mm}
\end{table*}
\subsection{Configurations}
This subsection provides the configuration details of the methods used in this study. All experiments were performed on a server with an Intel Core i7-10700K CPU, 32 GB of RAM, and an NVIDIA RTX 2080 Super GPU. Python $3.11.9$ and PyTorch $2.5.1$ were used for the implementation.
\begin{itemize}
\item \textbf{ILS.} The maximum number of iterations for optimizing each SFC is set to \( 10 \times |\text{SFC}| \times |\text{DCs}| \), following the settings recommended by the authors in~\cite{6226390}, ensuring sufficient exploration of the solution space per request.

\item \textbf{RAILS.} The configuration follows that of ILS, with the addition of risk models. The risk model consists of a two-layer MLP with $32$ neurons per layer, using the GELU activation function~\cite{hendrycks2017bridging}, Layer Normalization, and a residual connection. It maintains a memory size of $1000$ and is updated every $100$ steps using the AdamW optimizer with a learning rate of $0.01$. The acceptance probability threshold $\rho_{\text{risk}}$ is set to $0.5$ by default.

\item \textbf{paraDDQN and seqDDQN.} The Q-network constructed from two residual layers~\cite{he2016identity}. Each layer comprises two MLPs with a hidden dimension of $384$, employing Layer Normalization and GELU activation.
Sinusoidal positional encoding is utilized to encode positional information~\cite{vaswani2017attention}. The $\epsilon$-greedy strategy starts with $\epsilon = 1.0$ and decreases linearly by $0.1$ each episode until it reaches $0.1$. The replay buffer size is set to \( 10^6 \). The discount factor \( \gamma \) is $0.99$, and the learning rate is set to \( 10^{-4} \). A batch size of $256$ is used for training over $15$ episodes. AdamW optimizer~\cite{loshchilov2018decoupled} is adopted for weight updates.

\item \textbf{SDAC.} Both actor and critic networks consist of three Pre-Norm Transformer encoder layers. Each layer has a hidden dimension of $128$, $8$ attention heads, and an MLP dimension of $512$.
Sinusoidal positional encoding is employed to represent positional information.
The learning rate for the critic is set to \( 10^{-4} \), while the actor uses a lower learning rate of \( 10^{-5} \). The soft update parameter \( \tau \) for target networks is set to \( 10^{-3} \). The $\epsilon$-LoPe strategy starts with $\epsilon = 1.0$ and decreases linearly by $0.1$ at the end of each episode until it reaches $0.1$. The perturbation constant is empirically set to $c=0.1$.
Training is conducted with a batch size of $256$ over $15$ episodes using AdamW optimizer.
Note that to ensure a fair comparison, the number of parameters in the actor of SDAC is matched to that of paraDDQN and seqDDQN.
\end{itemize}
To assess the performance of the algorithms, a comprehensive evaluation is conducted. A total of $10$ substrate topologies are generated for the simulation environment using the \textit{Erdős–Rényi model}, where each edge is included with a probability of $0.5$.
The proposed ARN (see Section~\ref{sec:method}) is applied to all DRL-based approaches.
After training, the topology that yields the median average reward for each algorithm is selected as the test instance, ensuring a representative evaluation. For each selected topology, $10$ independent testing runs are performed using different random seeds to account for variability. The averages of the performance metrics obtained across these runs are reported in the subsequent section.

\section{Evaluation Metrics and Results}\label{sec:results}
This section presents an extensive performance evaluation of the proposed SDAC framework against SoTA (meta-)heuristic and DRL-based methods for SFCP problems. Through extensive simulations, we analyze key performance metrics, including long-term acceptance rates, causes of service request rejections, learning convergence, inference times, and resource utilization. Additionally, we conduct an ablation study on the proposed $\epsilon$-LoPe and ARN techniques, provide a computational complexity analysis of the framework, and discuss practical considerations for real-world deployment.

\subsection{Average Acceptance Rate}
The long-term acceptance rate is a critical performance metric that measures the proportion of SFC requests successfully accepted over the entire evaluation period, as described in~(\ref{eq:maximize_xy}). It reflects the effectiveness of a partitioning algorithm in managing system resources and meeting SLA requirements.

\begin{figure}[h]
    \centering
    \includegraphics[width=1\columnwidth]{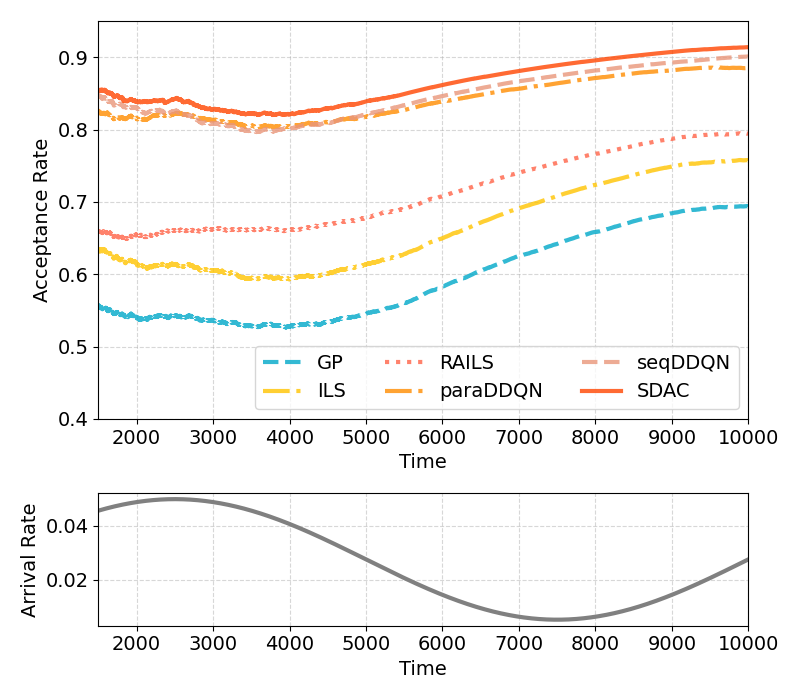}
    \caption{Acceptance and arrival rate over time.}
    \label{fig:acceptovertime}
    \vspace{-4mm}
\end{figure}

Fig.~\ref{fig:acceptovertime} presents the acceptance rate of different algorithms over time (top) and the corresponding arrival rate of SFC requests (bottom). In the early phase of the simulation, the acceptance rate is relatively low for all algorithms due to high demands, as the arrival rate reaches its peak during this period. As the arrival rate gradually decreases, the system experiences reduced traffic intensity, leading to an improvement in acceptance rates. Among the algorithms, the proposed SDAC consistently outperforms the others throughout the simulation.
In comparison, seqDDQN demonstrates competitive performance but lags behind SDAC, followed by paraDDQN. RAILS performs poorly in the earlier steps due to the immaturity of its risk models, which require more time to converge. Despite this initial drawback, RAILS eventually achieves better performance than the ILS, though it remains inferior to DRL-based approaches. The GP shows the lowest acceptance rate, highlighting its inefficiency in managing resources in general.
Overall, the results demonstrate that SDAC manages to sustain a higher acceptance rate even during high-traffic periods.
\begin{figure}[h]
     \centering
     \includegraphics[width=1\columnwidth]{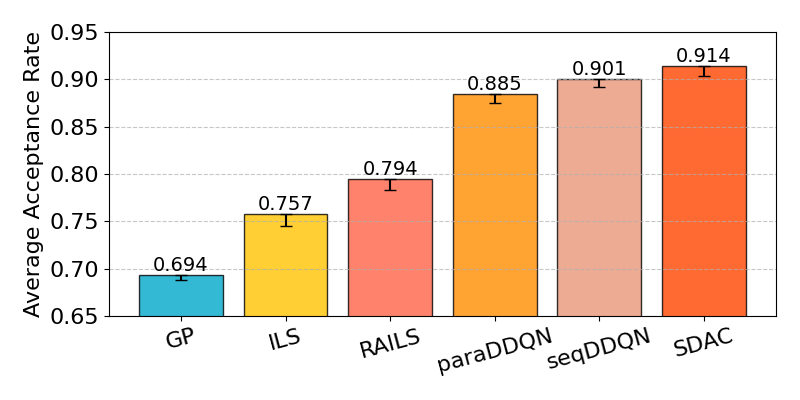}
    \caption{Average acceptance rate.}
    \label{fig:avgaccept}
\end{figure}
Fig.~\ref{fig:avgaccept} shows the average acceptance rate achieved by the different algorithms. SDAC exhibits the highest acceptance rate, while both seqDDQN and paraDDQN achieve competitive acceptance rates, although lower than SDAC. The stronger coordination of decisions in seqDDQN results in a performance advantage over paraDDQN. RAILS and ILS fall into the second tier, consistent with the results in Fig.~\ref{fig:acceptovertime}. GP again obtains the lowest acceptance rate.
To better understand the limitations of each approach, we further analyze the causes of service request rejection. The two dominant reasons are ($i$) insufficient CPU resources within DCs, leading to capacity violations, and ($ii$) violations of the E2E latency requirement specified in the SLA. Even when aggregated CPU capacity is available, fragmented or overloaded resources may prevent a request from being admitted, while in other cases the SFCs cannot be placed without satisfying the strict latency constraint.
\begin{table}[h!]
\centering
\caption{Causes of service request rejection.}
\begin{tabular}{|l|c|c|c|c|c|c|}
\hline
 \textbf{Violation} & \textbf{GP} & \textbf{ILS} & \textbf{RAILS} & \textbf{pDDQN} & \textbf{sDDQN} & \textbf{SDAC} \\
\hline
CPU (\%) & 21.8 & 21.0 & 13.1 & \textbf{5.0} & 8.8 & 8.1 \\ \hline
SLA (\%) &  8.4 &  3.3 &  9.5 & 5.6 & 0.6 & \textbf{0.3} \\
\hline
\end{tabular}
\label{tab:violations}
\end{table}
Table~\ref{tab:violations} shows the rejection percentages across algorithms. GP and ILS exhibit the highest overall rejection rates, with rejections dominated by CPU shortages around $21$\%, while SLA shares remain secondary. RAILS reduces CPU-related rejections compared to GP and ILS but still suffers from a notable share of SLA-driven failures at $9.5$\%. The DRL baselines reveal different trade-offs: paraDDQN lowers CPU rejections to $5.0$\% but accepts more SLA violations, whereas seqDDQN nearly eliminates SLA violations ($0.6$\%) at the cost of higher CPU rejection rate at $8.8$\%. SDAC attains the lowest rejection rate, nearly eliminating SLA violations ($0.3$\%) while keeping CPU-related rejections modest.

\subsection{Learning Curves and Inference Time}
We analyze the learning curves and inference time of the considered approaches. The learning curves depict the convergence behavior of the DRL-based models by showing the average testing rewards achieved over training episodes. Additionally, we evaluate the inference time per action, which measures the time taken by each method to make a single SFC partitioning decision during deployment.
\begin{figure}[h!]
     \centering
     \includegraphics[width=1\columnwidth]{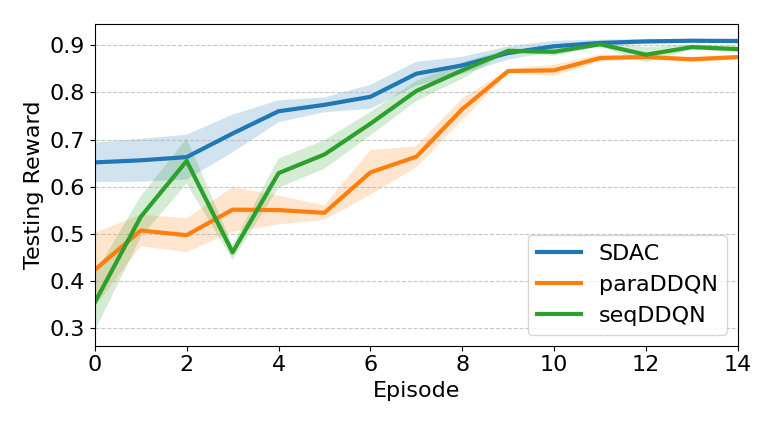}
    \caption{Average reward of DRL-based approaches over time.}
    \label{fig:lcurve}
\end{figure}
Fig.~\ref{fig:lcurve} presents the learning curves of the three DRL-based approaches. These curves reflect how efficiently and effectively each algorithm learns to optimize its policy over time.
All three methods exhibit steady improvement, indicating successful learning and convergence. SDAC starts with a relatively high reward and demonstrates the most stable and consistent growth throughout the training process. It also converges faster than the other methods and reaches the highest average reward.
seqDDQN also shows rapid learning, particularly after the initial episodes. Although it briefly dips in performance around episode $3$, it quickly recovers and nearly matches SDAC by the end of training.
paraDDQN, on the other hand, demonstrates slower and more gradual improvement. Its early-stage rewards lag behind the other two methods, and it converges to a slightly lower average reward. This suggests that while paraDDQN can learn effective policies, its parallel action structure may lead to less efficient coordination, resulting in comparatively slower convergence.
Overall, the figure highlights the effective learning dynamics of SDAC, which not only converges faster but also maintains high-reward progression across episodes. These advantages are particularly valuable in practical scenarios where computational resources or training time are limited.
\begin{figure}[h]
     \centering
     \includegraphics[width=1\columnwidth]{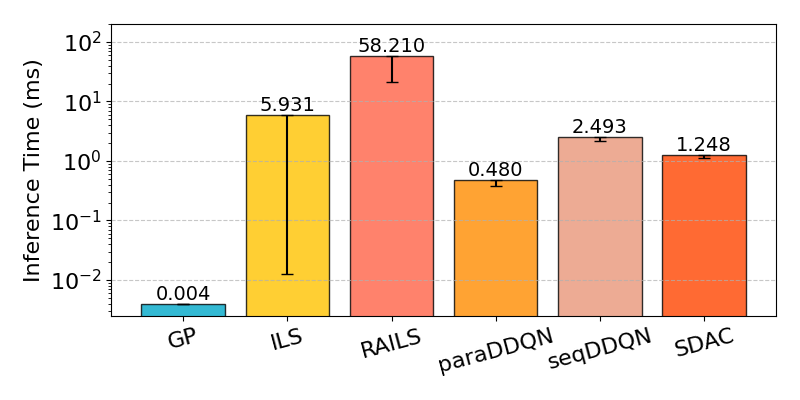}
    \caption{Average inference time per request.}
    \label{fig:infer}
\end{figure}
Fig.~\ref{fig:infer} presents the inference time per request on a logarithmic scale, highlighting significant differences in computational efficiency. GP is the fastest, with an inference time of $0.004$\,ms, reflecting its simplicity. ILS and RAILS exhibit the highest inference times, at $5.931$\,ms and $58.21$\,ms, respectively, due to their iterative optimization processes, with RAILS incurring additional overhead from handling risk models. Among DRL-based methods, paraDDQN achieves the lowest inference time of $0.480$\,ms, benefiting from parallel processing, while seqDDQN and SDAC are slightly slower but remain within a reasonable range.
Notably, when considered alongside the acceptance rates shown earlier in Fig.~\ref{fig:avgaccept}, SDAC achieves a strong balance between performance and efficiency. It maintains the highest acceptance rate while keeping inference time low.
Furthermore, the relatively small error margins for most methods indicate stable inference behavior. The only exceptions are ILS and RAILS, whose larger error bars reflect higher sensitivity to scenario variations due to their iterative optimization processes, which may terminate after an indefinite number of iterations depending on the complexity of the problem instances.

\subsection{Resource Utilization}
To evaluate the effectiveness of partitioning strategies, it is essential to analyze how computational resources are utilized across different DCs. Proper resource utilization not only ensures efficient operations but also reduces the likelihood of resource fragmentation, which can impact the acceptance rate of future SFC requests. This analysis considers two key metrics during peak hours, defined as the time periods with the highest demand for resources:
\begin{itemize}
    \item \textbf{Average Resource Utilization.}
    The average resource utilization metric provides an overall measure of how effectively the computational resources across all DCs are used during peak hours. For a set of DCs \( V_S \), the average resource utilization is defined as:
    \begin{equation}
    \frac{1}{|V_S|} \sum_{u \in V_S} \frac{\text{Resources Used at DC } u}{\text{Total Resources at DC } u}.        
    \end{equation}
    A higher value indicates better overall utilization of available resources, while a lower value suggests potential under-utilization.
    \item \textbf{Perplexity of Load Distribution.}
    The perplexity of resource utilization quantifies the distribution of resource usage across all DCs during peak hours. It reflects the degree of balance in resource allocation, where higher perplexity implies a more uniform distribution and lower perplexity indicates uneven utilization. For a set of DCs \( V_S \), the perplexity is computed as:
    \begin{equation}
    \exp\left(- \sum_{u \in V_S} p_u \log(p_u)\right),
    \end{equation}
    where \( p_u \) is the proportion of total utilized resources at DC \( u \), given by:
    \begin{equation}
    p_u = \frac{\text{Resources Used at DC } u}{\sum_{v \in V_S} \text{Resources Used at DC } v}.
    \end{equation}
    A higher perplexity value suggests that resource usage is evenly distributed across DCs, reducing the risk of overloading specific DCs, while a lower value indicates concentration of resource usage in a few DCs.
\end{itemize}
\begin{figure}[h!]
    \vspace{-4mm}
     \centering
     \includegraphics[width=1\columnwidth]{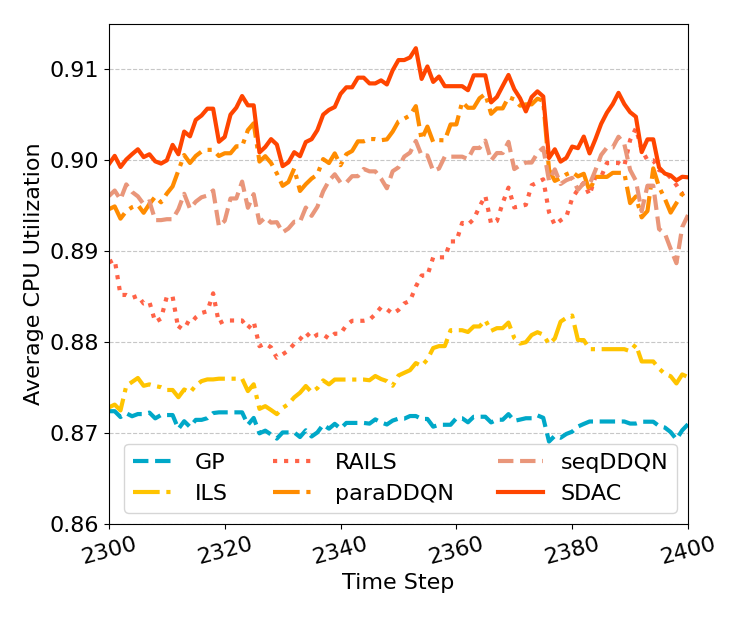}
    \caption{Average CPU utilization over peak hours.}
    \label{fig:cpuutil}
\end{figure}

Fig.~\ref{fig:cpuutil} shows the average resource utilization over time, which highlights the superior acceptance rate achieved by SDAC, with seqDDQN and paraDDQN following closely behind. This trend is consistent with the behavior observed in Fig.~\ref{fig:avgaccept}. As shown in the figure, SDAC consistently achieves the highest CPU utilization, peaking above 0.91. This indicates that SDAC makes the most effective use of available computational resources, which contributes directly to its high acceptance rates. Its ability to efficiently pack services onto infrastructure under heavy load reflects strong decision-making and orchestration capabilities.
seqDDQN and paraDDQN also exhibit high CPU utilization, though slightly lower than SDAC. seqDDQN maintains a slight advantage over paraDDQN across the time steps, consistent with its more coordinated decision-making framework. RAILS shows moderate utilization, gradually improving toward the end of the interval, but still trailing behind the DRL-based approaches.
On the other end of the spectrum, ILS and GP demonstrate the lowest CPU utilization. GP, in particular, maintains a consistently conservative resource usage pattern, suggesting a more cautious allocation strategy. While this may help avoid overloading the system, it results in significant underutilization of available resources and ultimately limits acceptance performance. Overall, the CPU utilization trends during peak hours strongly correlate with the algorithms’ effectiveness, confirming SDAC’s superior ability to adapt under pressure.
\begin{figure}[h!]
     \centering
     \includegraphics[width=1\columnwidth]{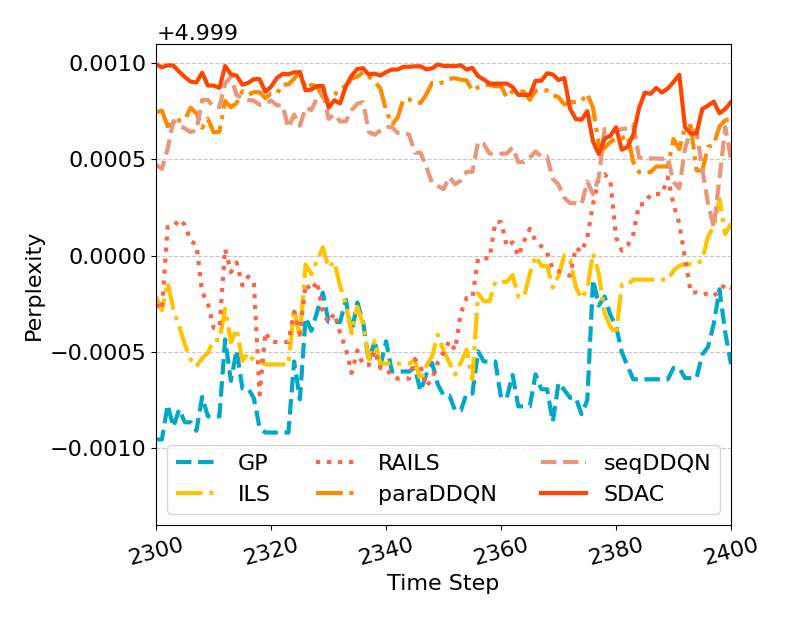}
    \caption{Perplexity of load distribution over peak hours.}
    \label{fig:loaddist}
\end{figure}

Fig.~\ref{fig:loaddist} illustrates the perplexity of load distribution over time, highlighting the balance of resource allocation across all DCs. The DRL-based methods maintain relatively high perplexity, indicating their ability to distribute resource usage evenly. In contrast, GP, ILS, and RAILS exhibit consistently lower perplexity, suggesting a more conservative approach that concentrates resource usage in fewer DCs. In particular, SDAC sustains high perplexity throughout the timeline and remains unaffected by noticeable drops observed in other algorithms, demonstrating its robustness in optimizing resource allocation over time.

\subsection{Ablation Study}
\label{ablation}
To assess the effectiveness of the proposed $\epsilon$-LoPe exploration strategy and ARN normalization technique described in Section~\ref{sec:method}, we conduct an ablation study with four different variants. Table~\ref{tab:ablation} summarizes the tested configurations that allow us to disentangle the effects of each component.
To ensure a fair baseline, we adopt Gumbel-Softmax~\cite{jang2017categorical} as the replacement for $\epsilon$-LoPe, providing a differentiable relaxation of categorical actions with stochastic sampling, and z-score normalization~\cite{JMLR:v22:20-1364} as the substitute for ARN, 
which represents conventional finite-horizon scaling method.
Comparing B against A isolates the benefit of $\epsilon$-LoPe, while C against A highlights the effect of ARN. Finally, D against A-C demonstrates the combined improvements achieved when both techniques are applied.
Note that to ensure a meaningful comparison, results are reported midway through training, since with sufficient training time, variants B and D eventually converge to similarly high performance levels, making full-convergence comparison less informative.
\begin{table}[h]
\centering
\caption{Configurations for the ablation study.}
\label{tab:ablation}
\begin{tabular}{|c|c|c|}
\hline
\textbf{Label} & \textbf{Exploration strategy} & \textbf{Normalization technique}\\
\hline
A & Gumbel-Softmax & z-score\\
\hline
B & \textbf{$\bm{\epsilon}$-LoPe} & z-score\\
\hline
C & Gumbel-Softmax & \textbf{ARN}\\
\hline
D & \textbf{$\bm{\epsilon}$-LoPe} & \textbf{ARN}\\
\hline
\end{tabular}
\end{table}
\begin{figure}[b]
     \centering
     \includegraphics[width=0.8\columnwidth]{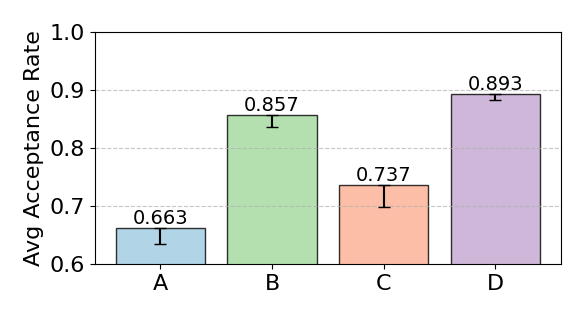}
    \caption{Average acceptance rates of variants listed in Table~\ref{tab:ablation}.}
    \label{fig:ablation_result}
\end{figure}

Fig.~\ref{fig:ablation_result} presents the average acceptance rates for the four ablation variants defined in Table~\ref{tab:ablation}. The baseline configuration A achieves the lowest acceptance rate, indicating that conventional stochastic exploration and normalization are insufficient to ensure effective policy updates. 
Introducing $\epsilon$-LoPe in B yields a clear improvement of $29.3\%$, showing that controlled logit perturbation promotes more efficient exploration. Variant C, which integrates ARN while retaining Gumbel-Softmax, attains a higher mean acceptance rate (an $11.2\%$ improvement over A) but also exhibits a larger variance. This suggests that ARN alone might mitigate gradient-scale instability yet remains sensitive to early reward fluctuations when exploration is not sufficiently diversified. 
The full configuration D, namely the proposed SDAC framework, achieves both the highest acceptance rate (a $34.7\%$ improvement over A) and the smallest variance, indicating that the two mechanisms are complementary: $\epsilon$-LoPe enhances exploratory robustness, while ARN regularizes long-term return magnitudes, jointly yielding a more stable and  effective learning process.
\subsection{Computational Complexity and Practical Considerations}
The computational complexity of SDAC can be analyzed for both inference (deployment) and training phases. Let $n = |V_v|$ denote the number of VNFs, $m = |V_s|$ the number of DCs, $L$ the number of Transformer encoder layers, $d$ the embedding dimension (model width), and $B$ the training mini-batch size. During inference, the actor network processes one SFC by passing the $n$ VNFs as tokens through $L$ Transformer layers. The self-attention operation scales as $\mathcal{O}(L n^2 d)$~\cite{vaswani2017attention}, since each token attends to all others, while the per-layer MLP block adds $\mathcal{O}(L n d^2)$, and the final decision head that scores each VNF against $m$ DCs contributes $\mathcal{O}(n m)$. 
Thus, inference is dominated by the quadratic $n^2$ term from attention, while dependence on the number of DCs is only linear. Training, by contrast, requires both forward and backward passes for actor and critic networks, leading to $\mathcal{O}\!\left(B(L n^2 d + L n d^2 + n m)\right)$ per update step, where $B$ stands for the batch size. In terms of scalability, increasing the chain length $n$ primarily stresses the $\mathcal{O}(n^2)$ attention mechanism, while adding more DCs $m$ only linearly increases the cost of the output layer.

From a deployment perspective, several challenges may arise. For example, operational networks must ensure strict  SLA compliance, whereas learning-based policies may occasionally produce unstable decisions, particularly under unseen conditions. In addition, integration with existing orchestration frameworks across multiple administrative domains may introduce extra engineering complexity and security concerns. Addressing these challenges will be essential to ensure robust and scalable deployment of learning-based SFC partitioning solutions in real-world 6G network environments.

\section{Conclusion and Future Work}
\label{sec:conclusion}
In this work, we proposed a novel Transformer-empowered Actor-Critic Reinforcement Learning framework for sequence-aware SFC partitioning in multi-domain network orchestration. By integrating Transformer encoders within an Actor-Critic architecture, our method leverages the self-attention mechanism to model inter-dependencies among VNFs, enabling coordinated and parallel decision-making. The introduction of the $\epsilon$-LoPe exploration strategy and ARN further enhances training stability and convergence, addressing the limitations of existing sequential and parallel VNF partitioning methods. 
Additionally, the innovative critic network, inspired by sentence encoders in LLMs, enables holistic evaluation of VNF assignments, capturing contextual relationships within SFCs.
Comprehensive evaluations demonstrate that our method outperforms SoTA methods, including both DRL-based and (meta-)heuristic-based approaches. 
The proposed method achieves higher long-term acceptance rates, improved resource utilization and scalability while maintaining efficient inference times. 
These results validate the effectiveness of our method in optimizing SFC partitioning under the complexities of multi-domain networks, making it a robust solution for beyond 5G and 6G network environments.
Future research directions include investigating a decentralized scheme for the proposed method to enhance scalability and robustness in multi-domain settings, which is expected to reduce latency, enhance fault tolerance, and effectively handle imperfect information inherent in multi-domain network environments.


\section*{Acknowledgment}
This research was partially funded by the HORIZON SNS JU DESIRE6G project (grant no. 101096466) and the Dutch 6G flagship project ``Future Network Services''.

\bibliographystyle{IEEEtran}
\bibliography{main}
\begin{IEEEbiography}
[{\includegraphics[width=1in,height=1.25in,clip,keepaspectratio]{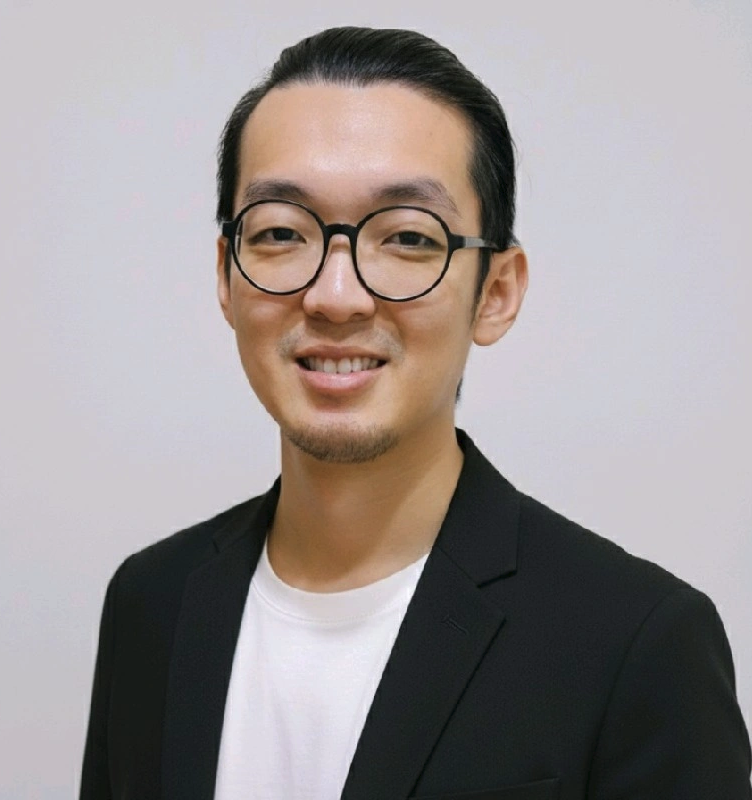}}]{Cyril Shih-Huan Hsu}
is a Postdoctoral Researcher at the Informatics Institute, University of Amsterdam, The Netherlands. He received his Ph.D. from the University of Amsterdam, Multiscale Networked Systems Group. He earned his B.Sc. and M.Sc. degrees from National Taiwan University in 2013 and 2015, respectively. From 2016 to 2021, he worked as a machine learning researcher at several international AI startups. His research focuses on leveraging AI and machine learning for efficient resource orchestration in complex networked systems, with an emphasis on multi-agent and decentralized environments.
\end{IEEEbiography}

\begin{IEEEbiography}[{\includegraphics[width=1in,height=1.25in,clip,keepaspectratio]{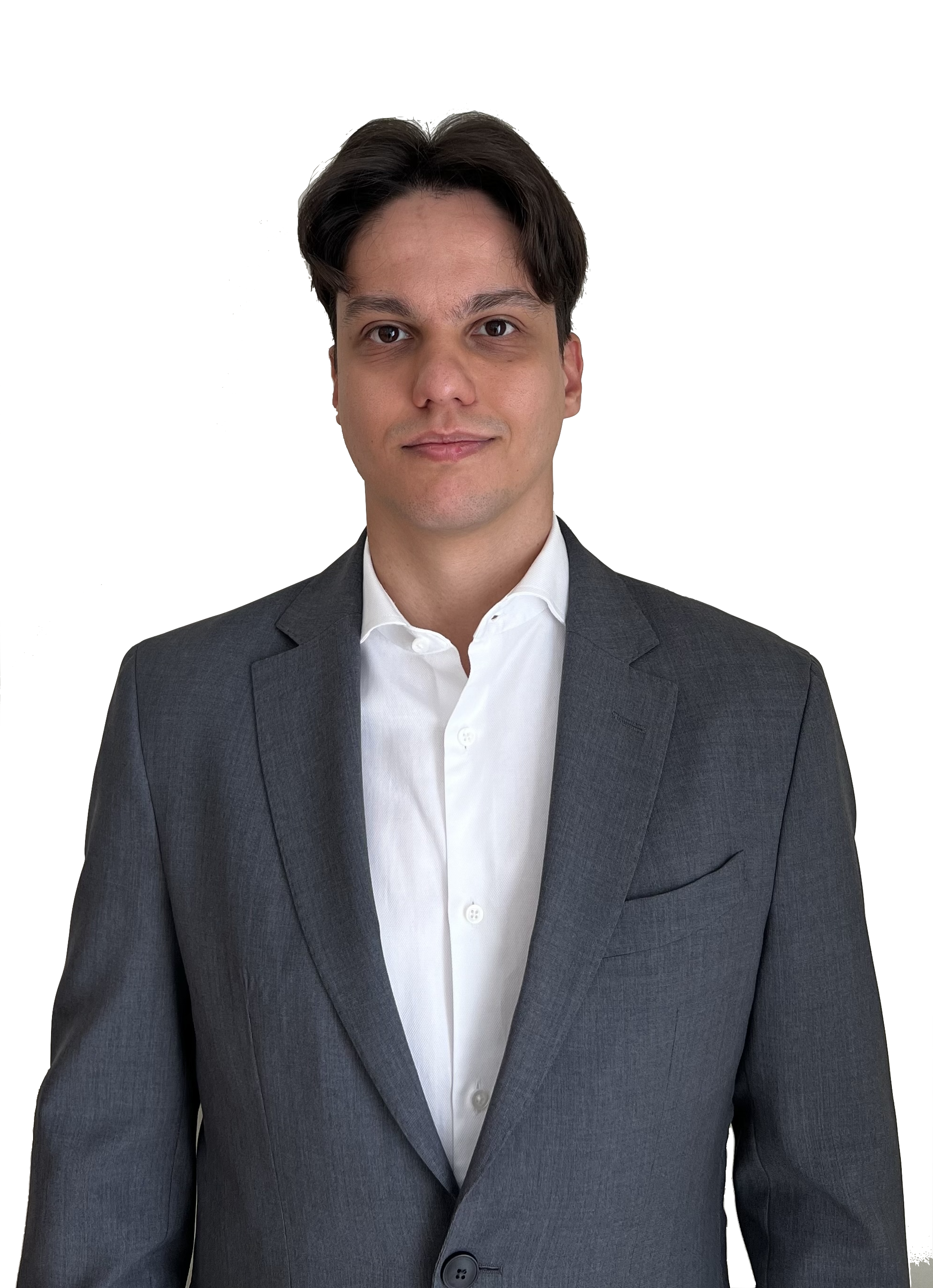}}]{Anestis Dalgkitsis} is a Postdoctoral Researcher at the University of Amsterdam’s Multi-Scale Networked Systems group. With a PhD in Signal Theory and Telecommunications from Universitat Politècnica de Catalunya, he specializes in decentralized network orchestration, AI-driven automation, and programmable data planes. Anestis has a strong background in 5G/6G research, multi-agent systems, and high-performance network experimentation, with his work earning recognition in international competitions and conferences. His career includes roles in prestigious EU projects and collaborations with global industry and academic partners.
\end{IEEEbiography}

\begin{IEEEbiography}[{\includegraphics[width=1in,height=1.25in,clip,keepaspectratio]{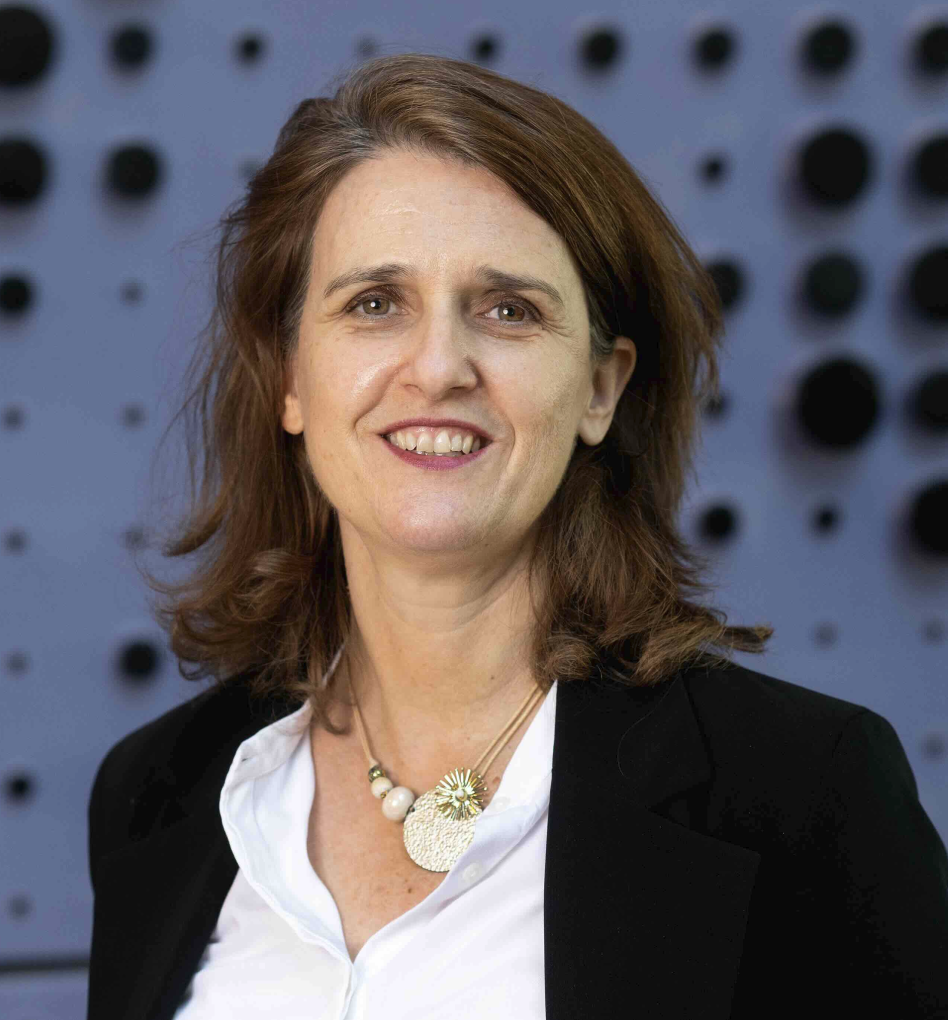}}]{Paola Grosso}
(Member, IEEE and ACM)  is a Full Professor at the University of Amsterdam with a chair on "Multiscale networks". She is part of the Multiscale Networked Systems research group (mns-research.nl) and director of the Informatics Institute (ivi.uva.nl). Her research focuses on the creation of sustainable and secure computing infrastructures, which rely on the provisioning and design of programmable networks. Current interests cover developments of control planes for quantum networks, in-band telemetry and use of FPGA for enhanced network monitoring and adoption of digital twins to support enhanced network operations. She has an extensive list of publications on these topics (https://www.uva.nl/profiel/g/r/p.grosso/p.grosso.html) and currently contributes to several national and international projects in this area.
\end{IEEEbiography}

\begin{IEEEbiography}[{\includegraphics[width=1in,height=1.25in,clip,keepaspectratio]{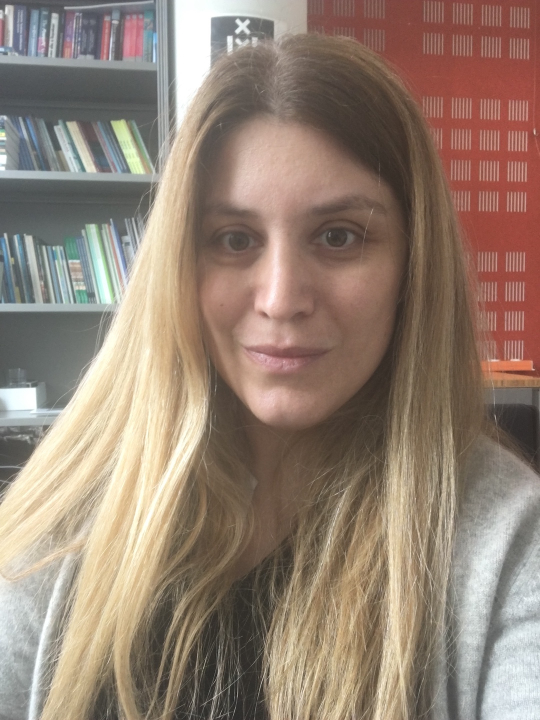}}]{Chrysa Papagianni}
is an Associate Professor at the Informatics Institute of the University of Amsterdam. She is part of the Multiscale Networked Systems group that focuses its research on network programmability and data-centric automation.  Her research interests lie primarily in the area of programmable networks with an emphasis on network optimization and the use of machine learning in networking. She has worked on multiple research projects funded by the European Commission, the Dutch research council and the European Space Agency, such as 5Growth, CATRIN, etc. Chrysa is currently coordinating the SNS JU DESIRE6G project on 6G system architecture. She also leads the AI-assisted networking work package in the 6G flagship project for the Netherlands on Future Network Services.
\end{IEEEbiography}


\end{document}